\documentclass[structabstract]{aa}  
\usepackage{graphicx,aalongtable}
\usepackage{amsmath}
\usepackage{txfonts}
\usepackage{subfigure}
\usepackage{lineno}
\usepackage{natbib}
\bibpunct[, ]{(}{)}{;}{a}{}{,}

\begin{document}
%
%\linenumbers
   \title{Anomalies in low-energy Gamma-Ray Burst spectra with the \textit{Fermi} Gamma-Ray Burst Monitor}

%   \subtitle{I. Overviewing the $\kappa$-mechanism}

   \author{D.~Tierney\inst{1},
          S.~McBreen\inst{1},
          R.~D.~Preece\inst{2},
          G.~Fitzpatrick\inst{1},
          S.~Foley\inst{1},
          S.~Guiriec\inst{3},
          E.~Bissaldi\inst{4},
          M.~S.~Briggs\inst{2},
          J.~M.~Burgess\inst{2},
          V.~Connaughton\inst{2},
          A.~Goldstein\inst{2},
          J.~Greiner\inst{5},
          D.~Gruber\inst{5},
          C.~Kouveliotou\inst{6},
          S.~McGlynn\inst{5,7},
          W.~S.~Paciesas\inst{8},
          V.~Pelassa\inst{2}
          and A.~von~Kienlin\inst{5}
          }

   \institute{University College Dublin, Belfield, Dublin 4, Dublin, Ireland\\
              \email{david.tierney@ucd.ie}
              \and
              Physics Department, University of Alabama in Huntsville, 320 Sparkman Drive, Huntsville, AL 35805, USA
              \and
              NASA Goddard Space Flight Center, Greenbelt, MD 20771, USA
              \and
              Institute for Astro and Particle Physics, University of Innsbruck, Technikerstrasse 25, 6020 Innsbruck, Austria
              \and
              Max-Planck-Institut f\"{u}r extraterrestrische Physik, Giessenbachstrasse 1, 85748 Garching, Germany
              \and
              Space Science Office, VP62, NASA/Marshall Space Flight Center, Huntsville, AL 35812, USA
              \and
              Exzellenz-Cluster Universe, Technische Universit\"{a}t M\"{u}nchen, Boltzmannstrasse 2, 85748 Garching, Germany
              \and
              Science and Technology Institute, Universities Space Research Association, 320 Sparkman Drive, Huntsville, AL 35805, USA
             }

%   \date{Received September 15, 1996; accepted March 16, 1997}

%\abstract{a}{b}{c}{d}{e}{f} 
 \abstract
   {A Band function has become the standard spectral function used to describe the prompt emission spectra of gamma-ray bursts (GRBs). However, deviations from this function have previously been observed in GRBs detected by BATSE and in individual GRBs from the \textit{Fermi} era.
}      
   {We present a systematic and rigorous search for spectral deviations from a Band function at low energies in a sample of the first two years of high fluence, long bursts detected by the \textit{Fermi} Gamma-Ray Burst Monitor (GBM). The sample contains 45 bursts with a fluence greater than 2$\times$10$^{-5}$ erg / cm$^{2}$ (10 - 1000 keV).   }
  {An extrapolated fit method is used to search for low-energy spectral anomalies, whereby a Band function is fit above a variable low-energy threshold and then the best fit function is extrapolated to lower energy data. Deviations are quantified by examining residuals derived from the extrapolated function and the data and their significance is determined via comprehensive simulations which account for the instrument response. This method was employed for both time-integrated burst spectra and time-resolved bins defined by a signal to noise ratio of 25 $\sigma$ and 50 $\sigma$. } 
   {Significant deviations are evident in 3 bursts (GRB\,081215A, GRB\,090424 and GRB\,090902B) in the time-integrated sample ($\sim$ 7\%) and 5 bursts (GRB\,090323, GRB\,090424, GRB\,090820, GRB\,090902B and GRB\,090926A) in the time-resolved sample ($\sim$ 11\%).} 
   {The advantage of the systematic, blind search analysis is that it can demonstrate the requirement for an additional spectral component without any prior knowledge of the nature of that extra component. Deviations are found in a large fraction of high fluence GRBs; fainter GRBs may not have sufficient statistics for deviations to be found using this method. 
   }
% 5 {} token are mandatory

  \titlerunning{Anomalies in low-energy GRB spectra with \textit{Fermi} GBM}
  \authorrunning{D. Tierney et al.}   
   
   \maketitle
%
%________________________________________________________________

\section{Introduction}
\label{intro}  
Gamma-ray bursts (GRBs) are the most luminous events in the universe and can briefly be summarised as having high-energy prompt emission followed by a multi-wavelength fading afterglow \citep[e.g.,][]{2009grbb.book.....V, 2011ApJ...734...96K}. The isotropic energy produced by a typical GRB is $E_{\rm iso}$ $\sim$ $10^{51}$ erg \citep[e.g.,][]{2001ApJ...562L..55F} and in some cases reaches up to $10^{54}$ erg \citep[e.g.,][]{2009A&A...498...89G,
2010A&A...516A..71M,2011ApJ...732...29C}. The prompt emission of GRBs has been detected over a wide spectral range from keV to GeV energies and is generally well modelled by one or a combination of the following: a smoothly broken power-law \citep[e.g.][]{1993ApJ...413..281B, 2009Sci...323.1688A}, a quasi-thermal component \citep[e.g.,][]{2000HEAD....5.3005P,2011ApJ...727L..33G,2011MNRAS.415.3693R,2012arXiv1207.6109T}, an extra non-thermal power-law component extending to high energies \citep[e.g.][]{2003Natur.424..749G,2008ApJ...677.1168K,2009ApJ...706L.138A} or a cut-off in the MeV regime \citep[e.g.,][]{2012ApJ...754..121T}. The lightcurves of GRBs are highly variable  and a number of studies into their temporal properties have been performed \citep[e.g.,][]{2002A&A...385..377Q,2011ApJ...740..104H,2012ApJ...744..141B}. 
 
Our work assumes that a Band function \citep{1993ApJ...413..281B} is the best fit function for GRB spectra in the GBM energy range and attempts to quantify the number of bursts that deviate from this function. In general, a Band function can be constrained better at lower energies compared to higher energies as more counts are observed at lower energies. This provides strict limits on the $\alpha$ parameter and makes it possible to search for deviations to the fit function. The Band model is defined as

\begin{equation}
N(E) = 
\begin{cases}
A \left( \tfrac{E}{100} \right )^{\alpha}\exp\left ( - \frac{E(2+\alpha)}{E_{peak}} \right ) & $if$ \quad E < E_{c}\\ 
A \left(\frac{(\alpha - \beta)E_{peak}}{100(2+\alpha)} \right)^{\alpha - \beta} \exp(\beta - \alpha) \left(\frac{E}{100}\right)^{\beta} & $if$ \quad E \geq E_{c}
\end{cases}
\end{equation}

where
\begin{equation}
E_{c} = (\alpha - \beta) \frac{E_{peak}}{2+\alpha}
\end{equation}

and \textit{A} is the amplitude in photons s$^{-1}$ cm$^{-2}$ keV$^{-1}$, $\alpha$ is the low-energy power-law index, $E_{\rm peak}$ is the $\nu$ F$_{\nu}$ peak energy in keV and $\beta$ is the high-energy power-law index.
\linebreak

The emission mechanisms for GRBs have generally been described as thermal or non-thermal with the thermal emission characterised by the radiation emitted from a cooling plasma \citep[e.g.][]{2000ApJ...530..292M}. However, thermal emission need not take the form of a standard Planckian model but may take a more complex form depending on several factors including the relative leptonic and hadronic populations, viewing angles and source region \citep[e.g.,][]{2011MNRAS.415.3693R}. Non-thermal emission can take the form of synchrotron emission from the extreme magnetic fields needed to create a GRB \citep[e.g.][]{2002ARA&A..40..137M}. Strong magnetic fields may also break and reconnect releasing energy in the form of gamma-ray photons \citep[e.g.][]{2011ApJ...726...90Z}. Although a Band function is not a physical model, it successfully models a large number of GRBs and it is useful to investigate the number of events which require additional paramaters. If an additional blackbody is observed, the temperature of the photosphere can be deduced \citep{2005ApJ...628..847R} whereas an underlying power-law can provide constraints on emission mechanisms \citep[e.g.][]{2009ApJ...705L.191A, 2010ApJ...709L.172R} and the Extra-galactic Background Light (EBL) \citep{2011ApJ...729..114A}. An overview of the observational and physical interpretations of the results from the \textit{Fermi} era are presented by \citet{2011BASI...39..471B}.

A number of papers discuss interesting bursts with unusual spectral behaviour in the keV range \citep[e.g.,][]{2011ApJ...727L..33G,2011MNRAS.415.3693R,2012arXiv1207.6109T}. It is important to note that there is a bias in the literature whereby interesting bursts are published more than bursts that conform to the standard GRB models. Another bias that can skew the apparent number of interesting bursts is that only bursts that look atypical initially are investigated further. A rigorous investigation must be carried out on a sample of GRBs in a systematic way to obtain the true fraction of GRBs with observable additional features. Any investigation must also be performed blindly to decrease the risk of interval selection effects which could bias the study. Here we present a systematic search for deviations from a Band function in the highest fluence bursts in the first catalogs from the \textit{Fermi} Gamma-ray Burst Monitor \citep{2012ApJS..199...18P,2012ApJS..199...19G}.

\subsection{\textit{Fermi} Gamma-ray Burst Monitor}

The \textit{Fermi} Gamma-ray Space Telescope, launched on 2008 June 11, has an energy range spanning several decades ($\sim$8 keV to  $\sim$ 300 GeV) and is ideal to explore the low-energy regime of GRBs. \textit{Fermi} consists of two instruments, the Large Area Telescope (LAT) operating between $\sim$~20 MeV to $\sim$ 300 GeV \citep{2009ApJ...697.1071A} and the Gamma-Ray Burst Monitor (GBM) operating between 8 keV - 40 MeV \citep{2009ApJ...702..791M}. 

GBM consists of two types of detectors - twelve Sodium Iodide (NaI) scintillating crystals operating between 8 - 1000 keV and two Bismuth Germanate (BGO) scintillating crystals operating between 0.15 - 40 MeV. The NaI detectors are arranged in clusters of three around the edges of the satellite and the BGOs are located on opposing sides of the satellite aligned perpendicular to the LAT boresight. As the detectors have no active shield and are uncollimated, GBM observes the entire unocculted sky. Further details can be found in \citet{2009ApJ...702..791M}.

Both NaI and BGO detectors collect counts in 4096 channels compressed into 128 energy channels, with boundaries spaced pseudo-logarithmically across the energy ranges of each detector. The lowest and highest energy channels in each detector are generally ignored in the analysis because of uncertainties in the instrument response. Extensive ground calibration was carried out on the detectors pre-launch \citep{2009ExA....24...47B} and in flight by comparison to other instruments  (INTEGRAL-ISGRI: \citet{2011AIPC.1358..397T}, INTEGRAL-SPI: \citet{2009AIPC.1133..446V} and \textit{Swift}-BAT: \citet{2009arXiv0907.3190S}). These calibration results are consistent and do not show any major changes from the on-ground calibration. Additional calibration work has also been carried out using the Crab Nebula with Earth occultation techniques \citep{2012arXiv1201.3585W}, nuclear lines from solar flares \citep{2012ApJ...748..151A}, positron/electron annihilation lines \citep{2011GeoRL..3802808B} and spectral analysis of Soft Gamma-ray Repeaters (SGRs) \citep{2012ApJ...756...54L}. 

\subsection{\textit{Fermi} GBM and BATSE/CGRO Comparison}
\label{GBMBATSE}
\begin{figure}
\centering
\includegraphics[width=9cm]{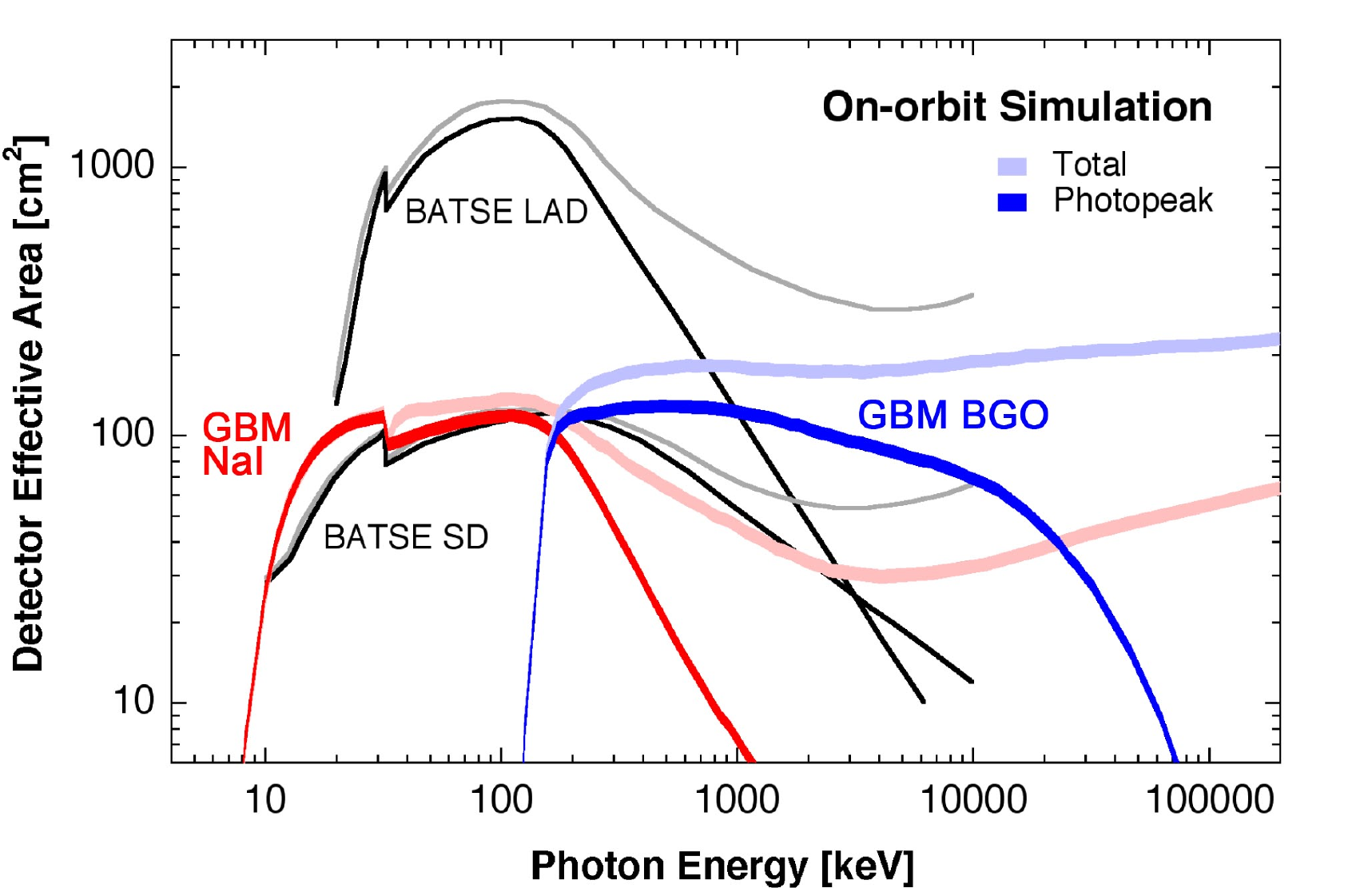}
   \caption{Plot comparing the effective areas of GBM v BATSE per detector. The total area exposed to a particular sky direction depends on the orientation of each detector and number of detectors exposed. At low energies the GBM NaI detectors have more effective area than the BATSE SDs and at high energies the GBM BGO detectors have more effective area that the BATSE LADs \citep{2009ApJ...702..791M}. The dark lines represent the photopeak effective area where the detected energy is the same (within the energy resolution) as the incident energy and the lighter lines represent the total effective area which includes the photopeak plus the cases where the instrument detects only part of the incident photon energy. (A color version of this figure is available in the online journal.)}
      \label{Eff_Area}
\end{figure}

A comprehensive study examining deviations at the lower end of the spectrum has not been carried out since \citet{1996ApJ...473..310P} investigated time-integrated GRB spectra using the spectroscopy detectors (SDs) on the Burst And Transient Source Experiment (BATSE) on board the Compton Gamma Ray Observatory (CGRO) \citep{1992ctap.conf..602G}. The results of this study showed that $\sim$~14\% of a sample of 86 GRBs contained significant low-energy excesses; no significant deficits were observed in the sample. A comparison between the effective areas of BATSE and GBM is given in Figure~\ref{Eff_Area}. The GBM NaI detectors are of similar specifications to the BATSE SDs (see Table~\ref{BATSEvGBM}). As the BATSE SD detectors were significantly thicker than the GBM NaI detectors, the BATSE SDs could detect higher energy photons than the GBM NaIs. This is compensated for on GBM by using the BGO detectors for high-energy constraints. 

The relative effective areas for each detector are presented in Figure~\ref{Eff_Area} showing that in the crucial energy range of interest (below $\sim$ 30 keV), the NaIs have a greater effective area than the SD detectors. Additionally, while \citet{1996ApJ...473..310P} only used a single SD detector, multiple NaI detectors were used in our analysis in all but 3 GRBs (see $\S$~\ref{SampleSelection}). GBM also has the advantage of greater effective area at higher energies using the BGOs. This provides better constraints on $\beta$ and $E_{peak}$ which help constrain the overall spectral model applied to the data. 

\begin{table}
\caption{Comparison of properties of the GBM NaIs with the BATSE SDs. }
%\begin{minipage}{\columnwidth}
\begin{tabular}{lll}
\hline
& BATSE SD\tablefootmark{1} & GBM NaI\tablefootmark{2} \\
\hline
Material & NaI & NaI \\
Number & 8 & 12 \\
Area & 126 cm$^{2}$ & 126 cm$^{2}$ \\
Thickness & 7.62 cm & 1.27 cm \\
Energy Range& 30 keV - 10 MeV\tablefootmark{*} & 8 keV - 1 MeV \\
Spectral Binning &  & \\
8 - 20 keV & 1 bin & $\sim$ 12 \\
\hline \\
\label{BATSEvGBM}
\end{tabular}
\tablefoot{(*) This energy range varied depending on the gain of the detectors where the lower end of the range could be as low as 5 keV for a high gain setting \citep{1996ApJ...473..310P}.} \\
\tablebib{(1)\citet{1996ApJ...473..310P}, (2) \citet{2009ApJ...702..791M} } \\
%\end{minipage}
\end{table}

To probe the lower energies using the BATSE SDs, two data types were required. The first data type was the Spectroscopy Discriminator data channels (DISCSP) which contained four integral channels over the entire energy range $\sim$ 5 keV - 2 MeV (for the highest gain setting). The upper edge of the lowest data channel (DISCSP 1) was set to the Lower-Level Discriminator (LLD) threshold of the Spectroscopy High-Energy Resolution Burst (SHERB) data ($\sim$ 10 keV for the highest gain setting). The SHERB data comprised 256 spectral energy channels between the LLD and $\sim$ 2 MeV. By using a joint fit between the SHERB data and the single DISCSP1 data point, \citet{1996ApJ...473..310P} determined the deviation of this single data point to the model. 
GBM has the advantage of a single standard data type from low to high energies. This reduces the chance of any incongruity between data sets. The data type also has better spectral resolution in the region of interest (see Table~\ref{BATSEvGBM}) so trends in the data can be more easily distinguished. GBM also has the advantage of better source locations provided by other instruments such as \textit{Swift} and the LAT which improves the ability to accurately model the instrument response. Out of the sample of GRBs analysed in this work, 47 \% have a sub-degree localisation.

%__________________________________________________________________

\section{Sample Selection}
\label{SampleSelection}
The sample was drawn from the first 2 years of triggered GBM GRBs (14$^{th}$ July 2008 - 13$^{th}$ July 2010). Bursts above a fluence of 2 $\times$ 10$^{-5}$ erg  cm$^{-2}$ (10 - 1000 keV) were  selected so that the spectral parameters could be constrained in the fitting process (see however $\S$~\ref{datacuts}). This gave a sample of 45 bursts, which formed the brightest 9\% of the 491 GRBs in the first GBM GRB catalog \citep{2012ApJS..199...18P}. In addition, 36 out of the 45 GRBs (80\%) have significant emission in at least one BGO detector as defined by \citet{2011ApJ...733...97B}. Only NaI detectors with source angles less than 60$^{\circ}$ were used in the spectral analysis to limit the effect of the uncertainties in the off-axis detector response. Two or more NaIs were used for 42 out of the 45 GRBs. The brightest BGO detector was used in all cases to constrain the spectral model at high energies.

As the signal to noise ratio (S/N) in these bursts is quite high, the influence of background fluctuations has a less statistically significant effect during the spectral fitting process compared to weaker bursts. Individual detectors were checked for blockages whereby the source photons passed through part of the spacecraft before entering a detector. This can cause photons to be scattered to different energies and lower energy photons to be absorbed beyond the ability of the detector response matrices (DRMs) to accurately reconstruct the photons incident on the entire spacecraft. An automated tool was initially used to check for line of sight blockages between the GRB and a detector. Further manual analysis was performed to remove any additional blocked detectors. Detectors that are blocked have a noticable deficiency of low-energy counts which is not consistent with other detectors that have similar source angles. Detectors which were not blocked and had an acceptable source angle were defined as `good' and used in the subsequent analysis.

\begin{figure*}
\centering
		\begin{subfigure}[]
                \centering
                \includegraphics[width=9cm]{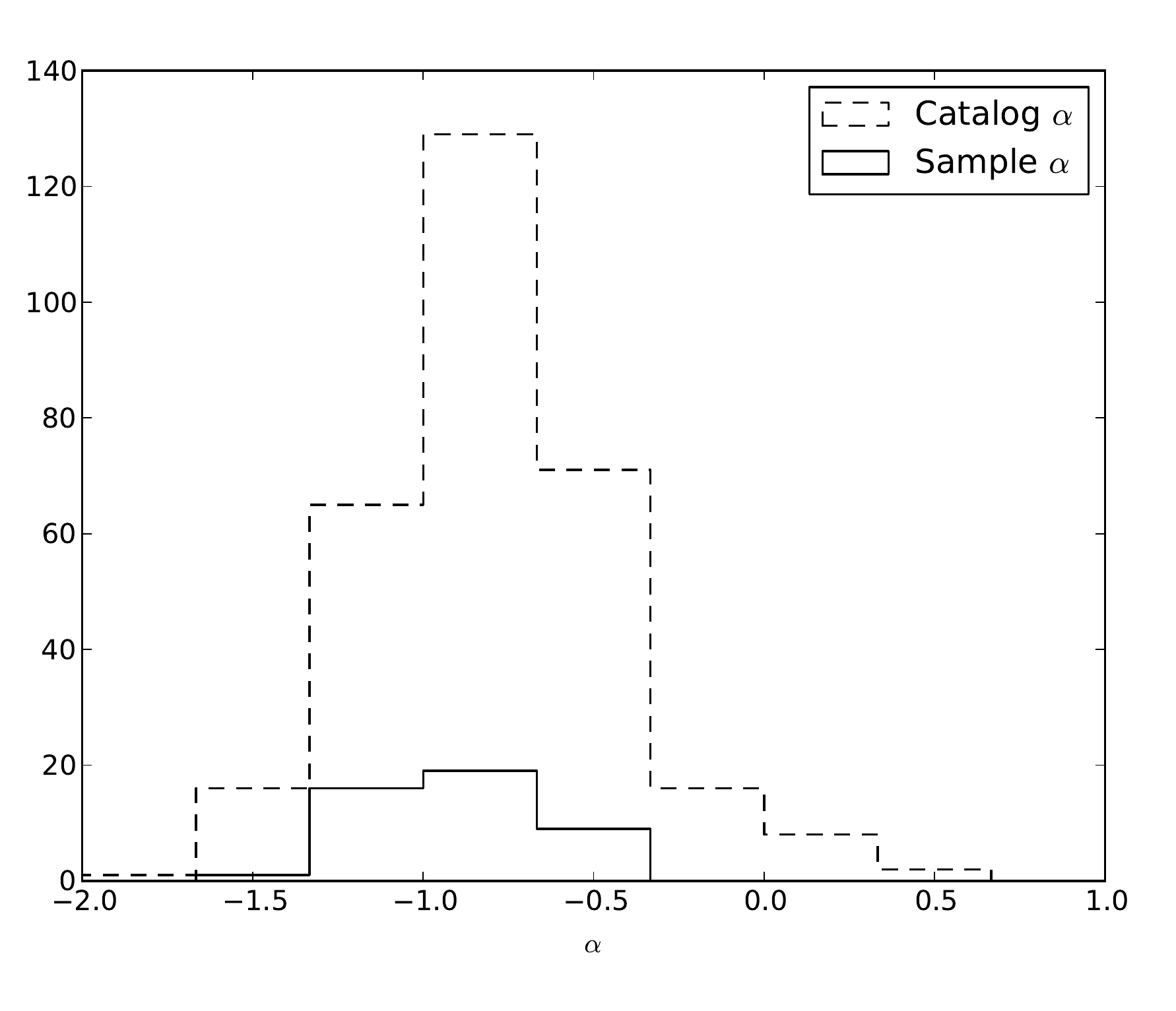}
        \end{subfigure}%
		\begin{subfigure}[]
                \centering
                \includegraphics[width=9cm]{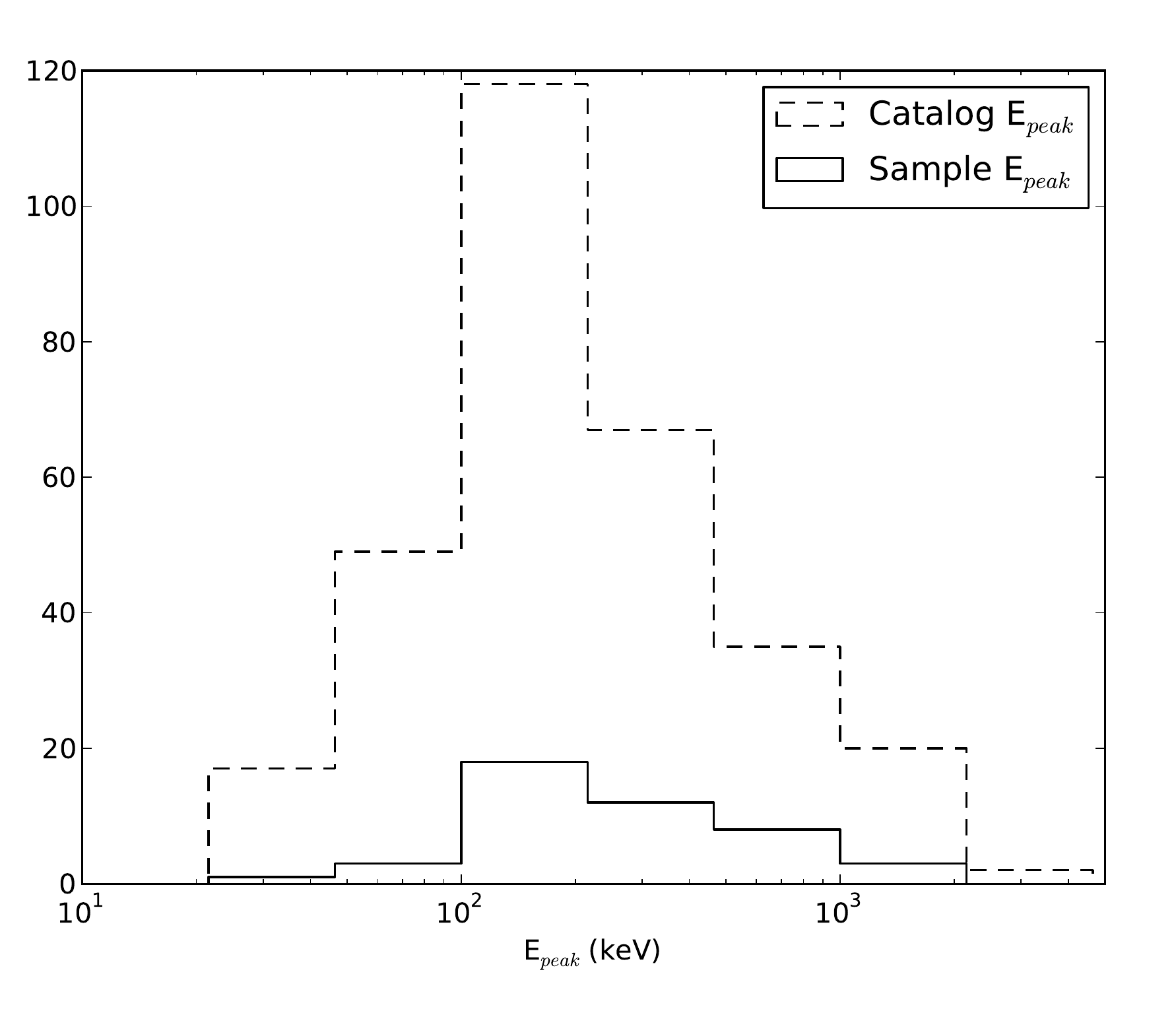}
        \end{subfigure}%
\caption{Histograms comparing \textbf{a)} $\alpha$ and \textbf{b)} $E_{\rm peak}$ in the high fluence sample from our work to a sample obtained from \citet{2012ApJS..199...19G}.} 
\label{AlphaVEpeak}
\end{figure*}

Figure \ref{AlphaVEpeak} shows the spectral properties of this sub-sample ($\alpha$ and $E_{\rm peak}$) relative to the overall GBM spectral catalog \citep{2012ApJS..199...19G}. The distributions show that the sub-sample selected here displays similar characteristics to the entire GBM spectral catalogue.

\noindent{}

\section{Method}
The sample was analysed using two methods. The first compares the data to the extrapolated fit of the function in the low-energy regime and the second compares the change in the spectral index $\alpha$ when the energy range of the data is shortened. These two methods are used in combination to identify deviations from a simple power-law at lower energies.

For an additional component to be significantly detected in the spectrum of a burst, it must either be continuously present throughout the entire burst or very strongly present in certain sections of the burst. An additional component may be present in certain sections of a GRB but not be intense enough to significantly alter the overall spectrum of the burst. Therefore the spectral fitting was performed on the spectrum of the entire burst (time-integrated) and on significant time slices (time-resolved) of the data. All spectral fitting was performed using the RMFIT software package \footnote{http://fermi.gsfc.nasa.gov/ssc/data/analysis/user/} and minor modifications were made to automate the fitting process. 

\subsection{Time-Integrated Single Fit}
CSPEC data were used for each `good' detector over the full energy range of GBM (8 keV - 40 MeV). These data have an energy resolution of 128 channels and a temporal resolution of 4 s pre-trigger changing to 1 s post-trigger until T0 + 600 s. Using the lightcurve, a time region encompassing the main emission phase of the GRB was selected by eye. A polynomial background was then fit to each GRB lightcurve by selecting background regions before and after the prompt phase. Multiple response matrices were used in the fitting process to account for spacecraft slewing relative to the source. In this process a new response is made when the spacecraft slews by more than 2 degrees, which can be important in long bursts, and is especially important if the spacecraft executes an autonomous repoint. A similar method was also employed by \citet{2012ApJS..199...19G}.

An initial time-integrated spectral fit was performed using a Band function for each GRB in the sample. Fit residuals were determined by the number of standard deviations that separate the data from a Band function. The residuals were summed between 8 keV and a variable Low-Energy Threshold (LET) in order to search for excesses or deficits. The LET was set at 15, 20, 25, 30, 50 and 100 keV.  However, this technique is not optimal as the fitting algorithm forces the function through any deviations that are present and minimises any deviations in any particular part of the spectrum. To reduce this issue, the LET was used as a lower energy bound for spectral fitting and the data points below the LET were compared to an extrapolated version of the function. This extrapolated fit is defined in the next section. 

Although the sample was defined to be composed of high fluence GRBs, this does not necessarily guarantee a large number of counts in all energy channels. Due to the potential low count rate / bin ratio, the fit (minimization) was performed using Castor statistics. This is similar to Cash-statistics \citep{1979ApJ...228..939C}. Castor statistics assume Poisson uncertainties per bin compared to $\chi^{2}$ statistics which assumes Gaussian uncertainties per bin. These are equivalent in the high count regime but diverge in the low count regime (less than $\sim$ 10 counts / bin). A similar minimisation method was used by \citet{2012ApJS..199...19G}.

\subsection{Time-Integrated Extrapolated Fit}
An improved  extrapolated fitting technique was devised such that a Band function was fit in the energy range from the variable LET to $\sim$ 40 MeV. The  function obtained from applying the model in this narrower range was then extrapolated to the lower energies in order to ascertain how well the function described the data below the LET. Deviations present between $\sim$ 8 keV and the LET are quantified by summing the residuals between $\sim$ 8 keV and the LET. 

\begin{figure*}
\centering
		\begin{subfigure}[]
                \centering
                \includegraphics[width = 6.8cm, angle=90]{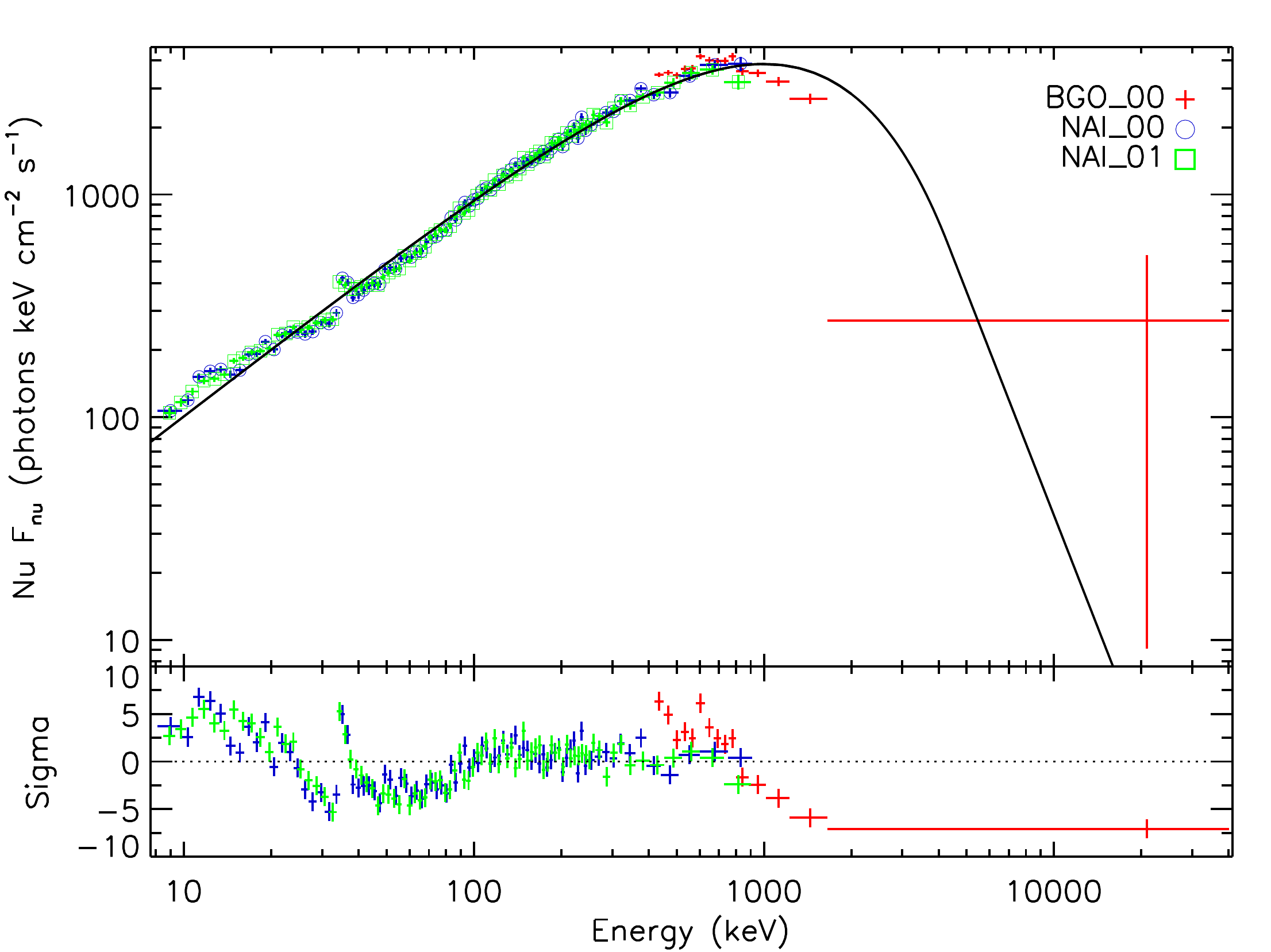}
        \end{subfigure}%
		\begin{subfigure}[]
                \centering
                \includegraphics[width = 6.8cm, angle=90]{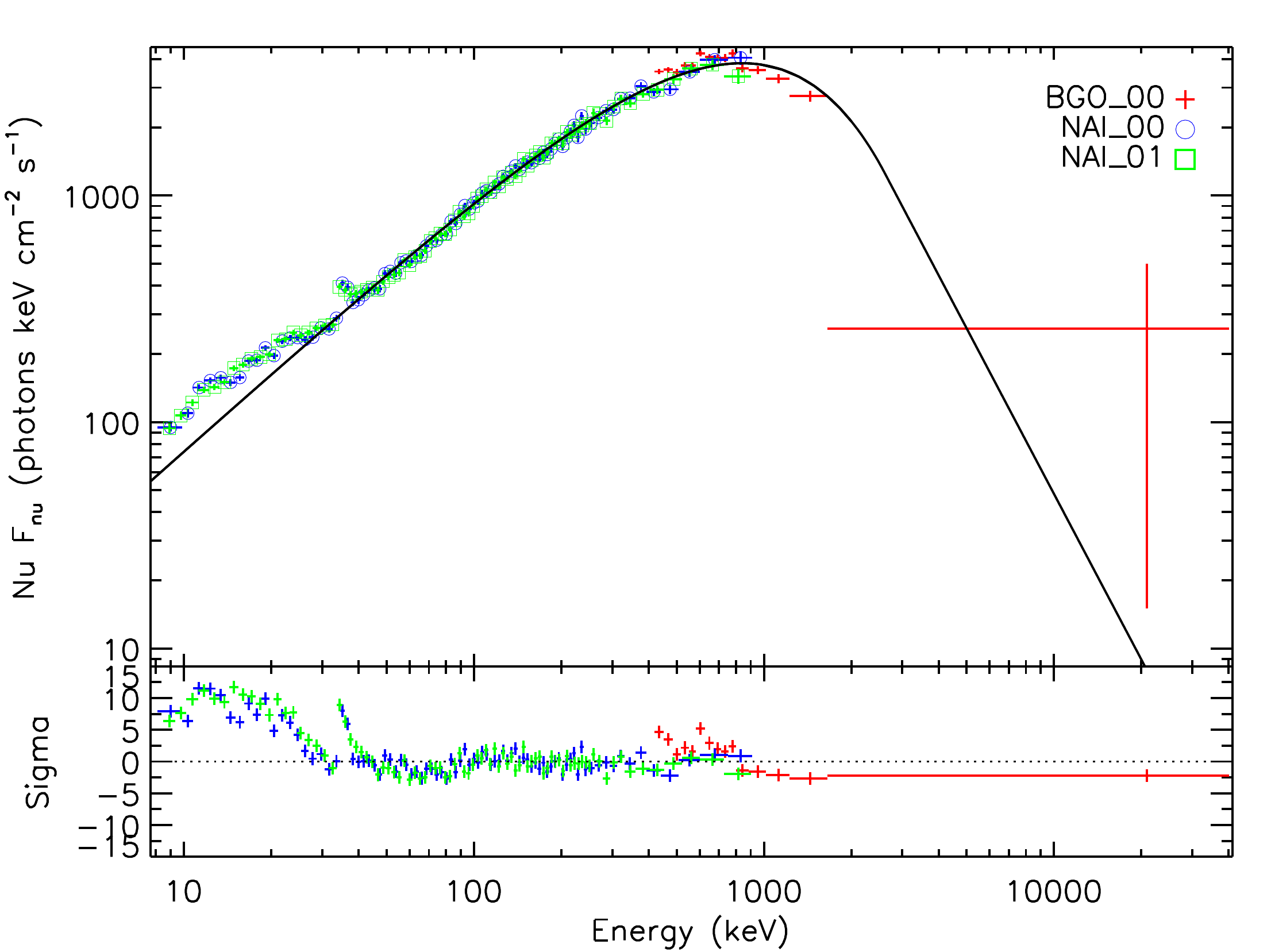}
        \end{subfigure}%
\caption{Comparing the single time-integrated fit to the extrapolated fitting technique. \textbf{a)} Single Band function fit to GRB\,090902B from 8 keV - 40 MeV ($\alpha$ = -0.99$^{+0.01}_{-0.01}$, $E_{\rm peak}$ = 996.8$^{+14.1}_{-14.2}$ , $\beta$ = -5.44$^{+0.89}_{-53.50}$). \textbf{b)} Band function fit to GRB\,090902B from 30 keV - 40 MeV and extrapolated down to 8 keV ($\alpha$ = -0.85$^{+0.01}_{-0.01}$, $E_{\rm peak}$ = 837.4$^{+11.9}_{-11.6}$ , $\beta$ = -4.42$^{+0.38}_{-0.71}$).} 
\label{extrapolatedfit}
\end{figure*}

One of the assumptions of the extrapolated fitting technique is that the function parameters obtained over a shorter energy range should be consistent with a fit performed over the entire energy range in the absence of any spectral excesses or deficits with respect to the function. The parameters obtained by fitting from 8 keV - 40 MeV should be consistent with those obtained by fitting from 15 keV - 40 MeV when no additional components are present. The difference between the single fit method and extrapolated fit method is shown in Figure \ref{extrapolatedfit}. Minimizing C-stat will tend to produce a fit that balances excesses and deficits regardless of the physical origin of the deviation from a Band function. In the single fit method, the deviation at low energies results in strong fluctuations throughout the entire spectral energy range. When the extrapolated fit method is used, the spectral deviations are observed more clearly at lower energies and the effects on the remaining spectrum are reduced. Features in certain energy bands are not physically comparable between different GRBs because the energy bands are defined in the observer frame and redshifts are not known for all GRBs. However, when performing an analysis on the manifestation of observable deviations in certain energy bands, fits using the same LET can be compared using this method.

The low-energy power-law index, $\alpha$,  must be reasonably well constrained in order to confidently calculate the significance of any deviations. This implies that $E_{\rm peak}$ and the LET must be sufficiently different to ensure that $\alpha$ can be constrained over the shortened energy range. The technique also requires that $E_{\rm peak}$ is greater than the LET. 

For each GRB, the spectral residuals between 8 keV and the LET are summed in an individual detector. The overall value for a particular GRB is the arithmetic mean for deviations in all detectors used in the analysis of the burst.  The Iodine K-edge region of the spectrum around 33 keV can lead to larger residuals caused by systematics rather than a process intrinsic to the source. However, a strong K-edge effect was only noted in $\sim$ 5 GRBs and dominated only 2 spectral bins. If the K-edge is present between 8 keV and the LET, it only weakly affects the summing of residuals over this range.  In reality it is only relevant when LET = 50 or 100 keV and the effect was not deemed significant enough to alter the method.

\noindent{}

\subsection{Time-Resolved Extrapolated Fit}
The spectrum of a GRB can evolve over the prompt emission interval \citep[e.g.][]{2000ApJS..126...19P} and a time-resolved analysis is required to account for such evolution. An S/N approach was employed over a basic time-resolved analysis (i.e. splitting the bursts into time sections) to ensure significant counts per bin. The selected region for each GRB used in the time-integrated fitting (full burst) was binned to the $25 \sigma$ and $50 \sigma$ level above background in the brightest detector. These intervals were then used to define bins for all other detectors in a particular GRB. A fit was then performed on each newly defined bin. 

\subsection{Simulations}

\label{simulations}
The goodness of fit of the data below the LET and the function fit at higher energies cannot easily be evaluated using C-Stat as there is no analytic result available to convert from the fit statistic to a measure of goodness of fit. Therefore it is difficult to quantify the significance of a deviation or even an acceptable range of values which occur in the absence of an excess/deficit at low energies. Simulations are therefore necessary to quantify whether a deviation is consistent with the extrapolated fit. 

A boot-strapping method was used to compare the actual results from the data with simulated results. In order to perform the simulations, the background and source must be simulated. The background was simulated using a similar level to that in the real data. To obtain the source region, a perfect Band function specific to each GRB, or time interval selected, was simulated. Multiple count distributions were created using this function (varying by Poisson noise). Each set of parameters was simulated $\sim$ 1000 times. 

For each simulated spectrum, a distribution of low-energy deviations using different LETs can be compiled. The simulated distributions were then fit with a Gaussian distribution to obtain a value for the mean and standard deviation (see Figure \ref{Null_Sim}). The data obtained by summing the low-energy residuals below the LET were then normalised by calculating the number of standard deviations by which the data varied from the mean of the simulated distribution. If the data for a GRB differ significantly from its simulated distribution, then it is claimed that a low-energy deviation is present.  

\begin{figure*}
\centering
		\begin{subfigure}[]
                \centering
                \includegraphics[width=9cm]{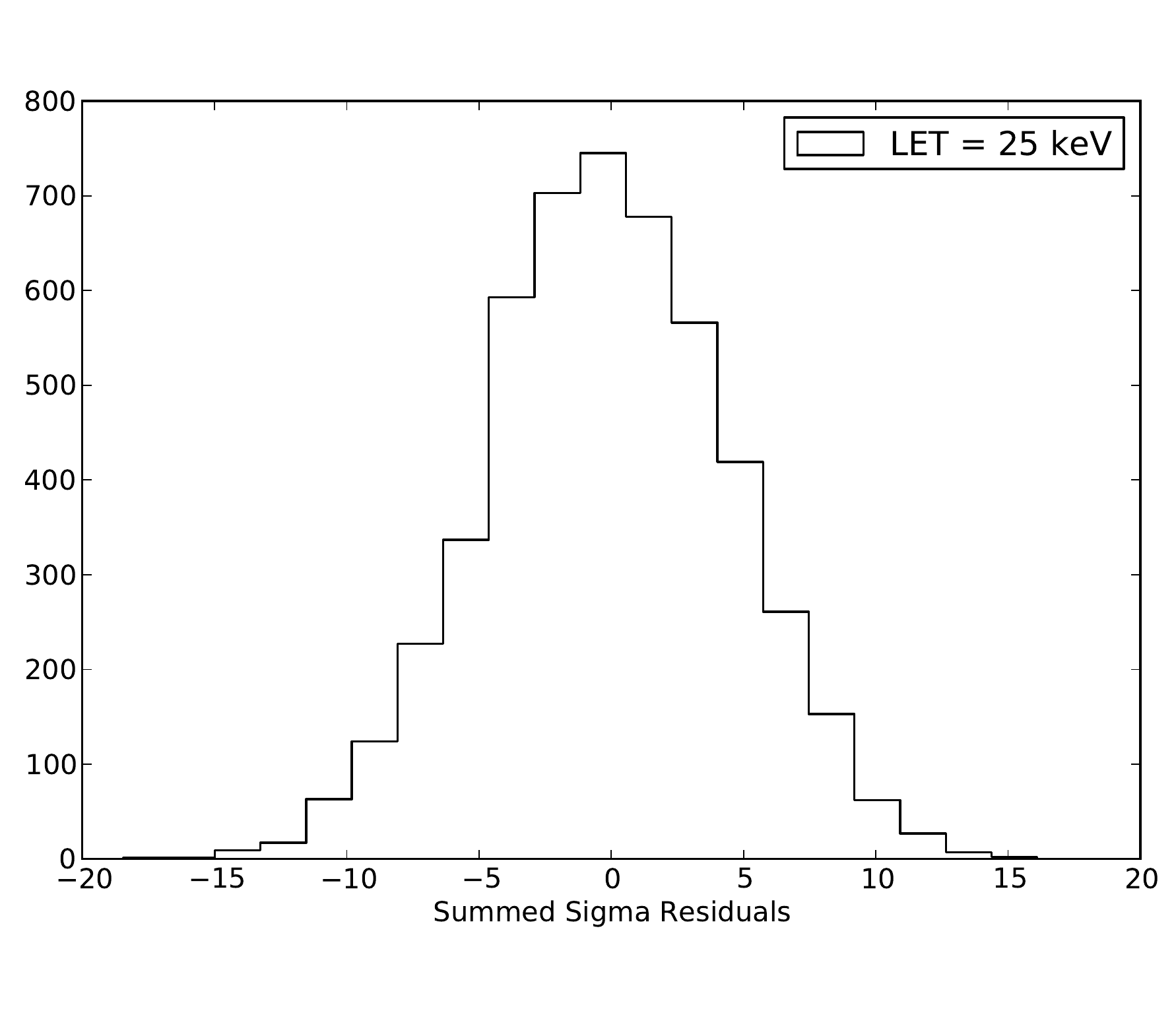}
        \end{subfigure}%		
		\begin{subfigure}[]
                \centering
                \includegraphics[width=9cm]{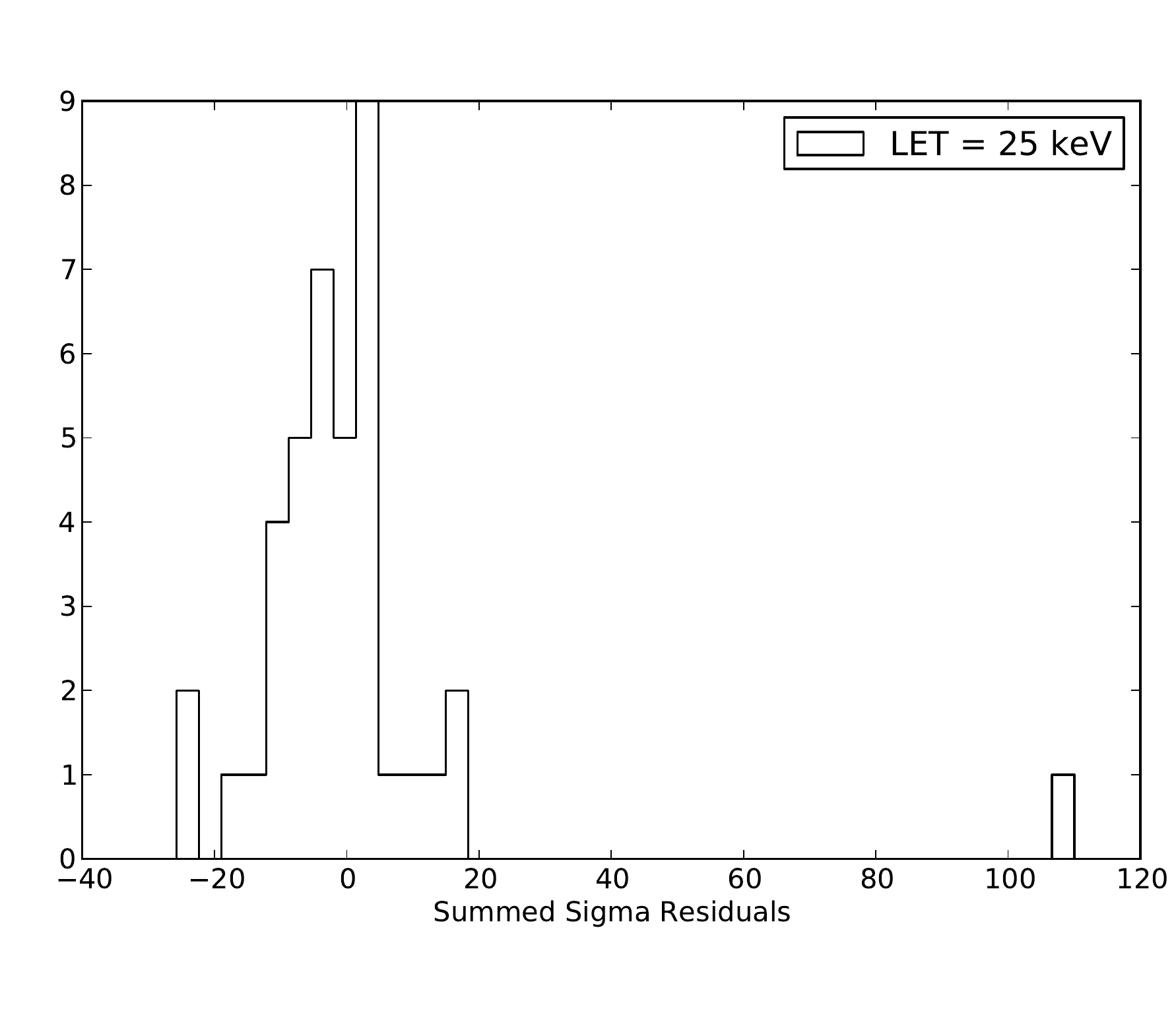}
        \end{subfigure}%
\caption{Comparing the time-integrated data to a simulated distribution. \textbf{a)} Combined distribution of low-energy deviations of 5 individual GRBs assuming a perfect Band function (no deviations present). The extrapolated fit method was used with LET = 25 keV. \textbf{b)} The low-energy residuals of the GRBs in the sample that survived the data cuts (see $\S$~\ref{datacuts}) for LET = 25 keV. The data distribution \textbf{b)} is broadly similar to the simulated distribution \textbf{a)}. The data point outside the main data distribution (on the right) is GRB\,090902B.}
\label{5SimGRBs}
\end{figure*}

A representation of the overall distribution was simulated to provide an initial insight into the expected distribution of residuals assuming that all spectra were correctly described by a Band function. Five GRBs were selected from the sample which represented a broad range of peak energies, with $E_{\rm peak}$ = 161, 276, 444, 484 and 1010 keV. These GRBs were simulated and the resulting distribution of residuals in the absence of excesses/deficits are presented in Figure \ref{5SimGRBs}. Individual simulations were then performed on a range of time-integrated and time-resolved sections of GRBs.

\noindent{}

\subsection{Data Cuts}
\label{datacuts}
Several cuts were applied to the data before analysis. If any fit failed in the automated fitting process due to a lack of spectral constraints, it was excluded from the sample. A range of initial parameters were passed to the fitting process in order to obtain a satisfactory solution, but 9 fits failed in the time-resolved fitting process despite alternative initial parameters and were excluded from the sample. These fits were individually examined and the failure was usually due to low $E_{\rm peak}$ and a lack of counts above $E_{\rm peak}$. No fits failed in the time-integrated analysis.

As the low-energy spectral index $\alpha$ only approaches a power-law in the asymptotic limit, a sufficient data range is needed to obtain reasonable constraints on the index. If $E_{\rm peak}$ is too close to the LET, two issues may arise 1) $\alpha$ will not approach the limit and 2) there will not be sufficient data to constrain $\alpha$ resulting in large fluctuations (e.g. LET = 50 keV, $E_{\rm peak}$ = 60 keV).  A well constrained $\alpha$ parameter is required because slight variations in $\alpha$ can cause large spectral deviations when extrapolated to lower energies.
For a spectrum to be accepted in the sample the following criteria must be met:
\begin{itemize}
\item  For LET = 15, 20, 25, 30 keV,  $E_{peak}$ $>$ 100 keV is required
\item  For LETs = 50 or 100 keV,  $E_{peak}$ $>$ 200 keV is required
\item  All intervals with a large proportional error on $E_{\rm peak}$, $\Delta$ $E_{\rm peak}$/$E_{\rm peak}$ $>$ 0.45 were excluded.
\item  All intervals with an error on $\alpha$ $>$ 0.2 were discarded to ensure that $\alpha$ can be reasonably extrapolated beyond the range of the fit.
\end{itemize}
The final sample after the cuts contains GRBs and intervals with good statistics and well constrained spectral parameters. Table \ref{Cuts} shows the number of GRB intervals both before and after the cuts were applied for the time-integrated and time-resolved analysis.

\begin{table}
\caption{Number of sample GRBs / GRB intervals pre- and post-cuts, where TI are the time-integrated results and 25$\sigma$/50$\sigma$ are the time-resolved results.}
\begin{tabular}{lllllll}
\hline
LET & TI  & TI & 25$\sigma$ & 25$\sigma$ & 50$\sigma$ & 50$\sigma$\\
 & Pre & Post & Pre & Post & Pre & Post\\
\hline
15 & 45 & 42 & 672 & 457 & 309 & 230\\
20 & 45 & 41 & 672 & 438 & 309 & 227\\
25 & 45 & 41 & 672 & 389 & 309 & 222\\
30 & 45 & 41 & 672 & 338 & 309 & 217\\
50 & 45 & 23 & 672 & 189 & 309 & 117\\
100 & 45 & 22 & 672 & 81 & 309 & 83\\
\hline
\end{tabular}
\label{Cuts}
\end{table}

\subsection{Variance of $\alpha$}
Another indication that a time interval in a burst had a low-energy spectral deviation was a variation in $\alpha$ when only the LET is changed. For each time section, each $\alpha$ parameter obtained from using different LETs was compared. If the value of $\alpha$ remains consistent across all different LETs in a spectrum, it is assumed that no significant low-energy deviations are present. If variation in $\alpha$ of $>$ 2 $\sigma$ for different LETs was evident, the GRB was investigated further. Usually a burst with a strong deviation in $\alpha$ would vary across multiple different LETs. If an interval had a variation in $\alpha$ and there was a deviation present in the low-energy residuals, this interval was concluded to have a significant spectral deviation from a Band function. All GRBs that had large deviations in the low-energy residuals displayed the expected variations in $\alpha$. 

\subsection{Summary of Method}
A brief summary of the method for time-resolved analysis is presented below.
\begin{enumerate}
\item The highest fluence GRBs from the first 2 years were selected for analysis. 
\item For each burst, NaI detectors $<$ 60$^{\circ}$ to the source and without blockages were selected. Multiple NaIs and a single BGO were selected in most cases.
\item Background and source regions were selected for an individual burst.
\item The main emission (time-integrated) interval of the GRB was selected by eye.
\item The GRBs were binned by S/N over the time-integrated interval of the GRB to produce the time-resolved intervals. 
\item The individually significant S/N bins were then fit with a Band function from a Low-Energy Threshold (LET) to $\sim$ 40 MeV. 
\item The LETs were selected to be 15, 20, 25, 30, 50, 100 keV. 
\item The fit function was then extrapolated to 8 keV. 
\item The residuals were summed between 8 keV and the LET for each detector and averaged across detectors. 
\item Simulations were then run and fit with a Gaussian distribution.
\item The significance of deviations in the data were then quantified by the number of standard deviations by which the data varied from the mean of the simulated distribution.
\item The $\alpha$ parameters between different LETs for a time region were compared. 
\item If a time region had a significant residual deviation and a variation in $\alpha$ while only changing the LET, it was considered to be a spectral deviation. 
\end{enumerate}

\section{Results}
In the context of this work, spectral deviations can be classified into two broad categories - excesses and deficits. The former occur when there is an excess of photons with respect to the extrapolated fit function below the LET. These, for example, could be explained by an additional component such as a power-law or blackbody dominating at low energies. Deficits occur, not due to a lack of photons at low energies but because $\alpha$ 
is flatter than the actual data at low energies. This can be caused by the presence of an additional component, such as a blackbody, between the LET and $E_{\rm peak}$ resulting in a shallower $\alpha$ and a deficit of photons relative to the extrapolated function at low energies.

\subsection{Hypothesis Testing}
\begin{figure*}
\centering
		\begin{subfigure}[]
                \centering
                \includegraphics[width=9cm]{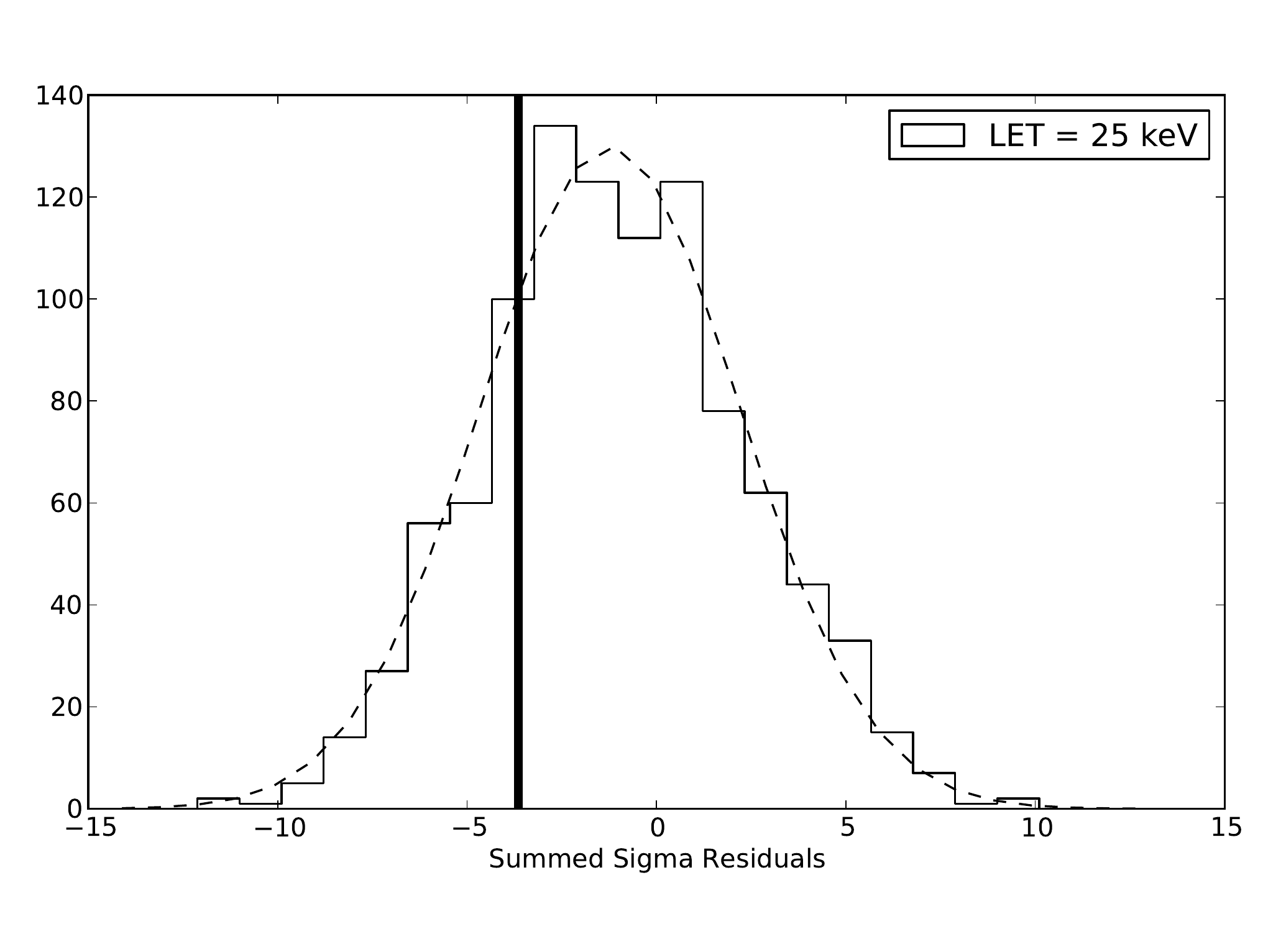}
        \end{subfigure}%
		\begin{subfigure}[]
                \centering
                \includegraphics[width=9cm]{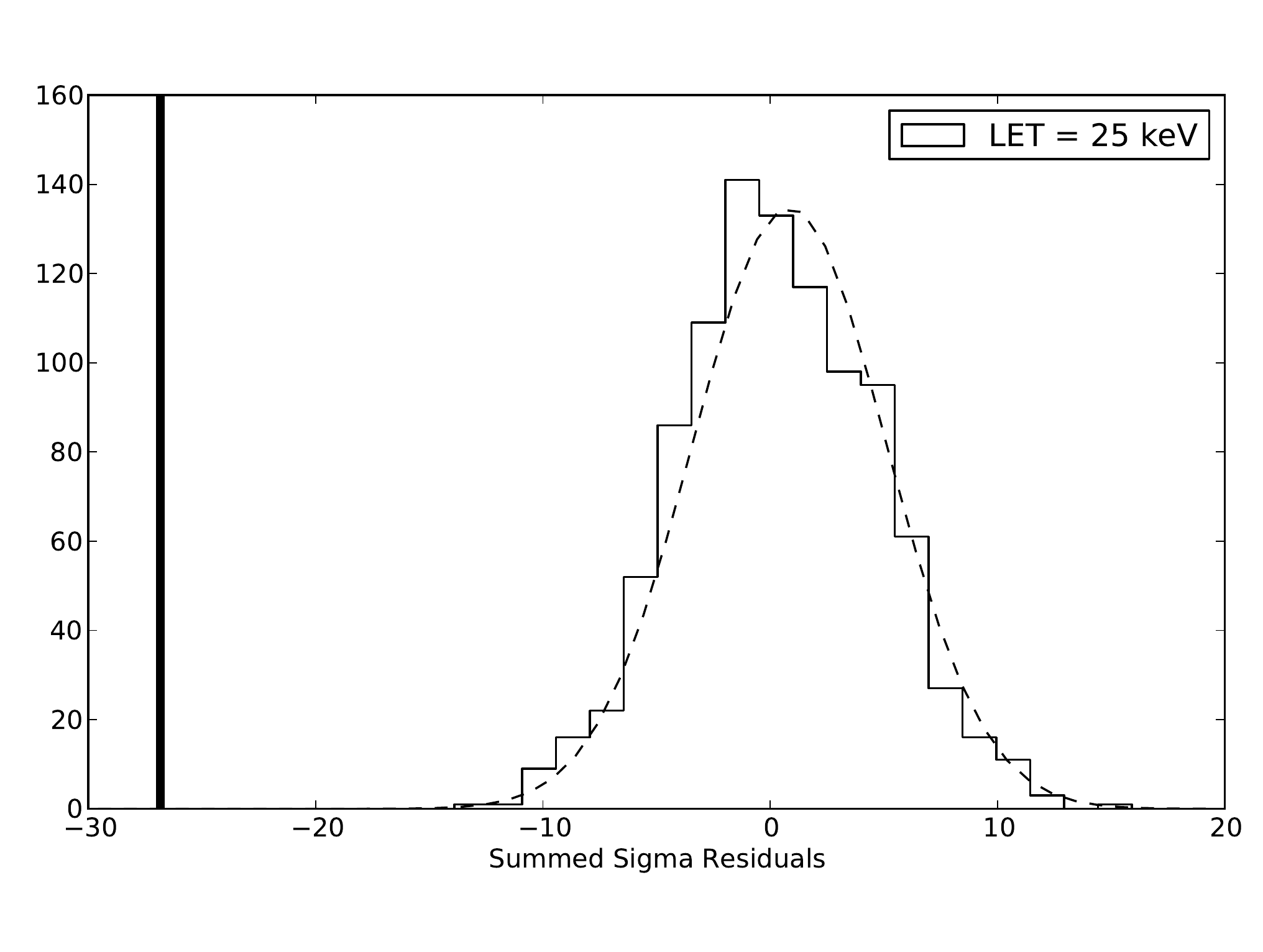}
        \end{subfigure}%
        \caption{Histograms of summed residuals below 25 keV obtained from time-integrated simulations of a perfect Band function. The dark line shows the value of the data for that GRB. \textbf{a)} The value of the deviation for GRB\,080817A is -3.6. The mean and the standard deviation of the simulated distribution are -1.1 and 3.4 respectively. In this case the data does not deviate significantly from the simulated distribution and has a normalised deviation of -0.7 $\sigma$. \textbf{b)} The value of the deviation for GRB\,090424 is -25.8 compared to the simulated distribution with a mean and standard deviation of 0.4 and 4.3 respectively. The data deviates significantly from the simulated distribution with a normalised deviation of -6.1 $\sigma$.}
\label{Null_Sim}
\end{figure*}

Two methods were used to test for deviations at low energies.  The first involves comparing the residuals from the data to those from simulations where a Band function is the assumed best fit. The results in Figure \ref{Null_Sim} show two GRBs, one with and one without deviations using the time-integrated results. The null hypothesis for these simulations is that if the data are consistent with the simulated distribution, the extrapolated spectrum is consistent with a Band function.

\noindent{}

\begin{figure*}
\centering
		\begin{subfigure}[]
                \centering
                \includegraphics[width = 6.8cm, angle=90]{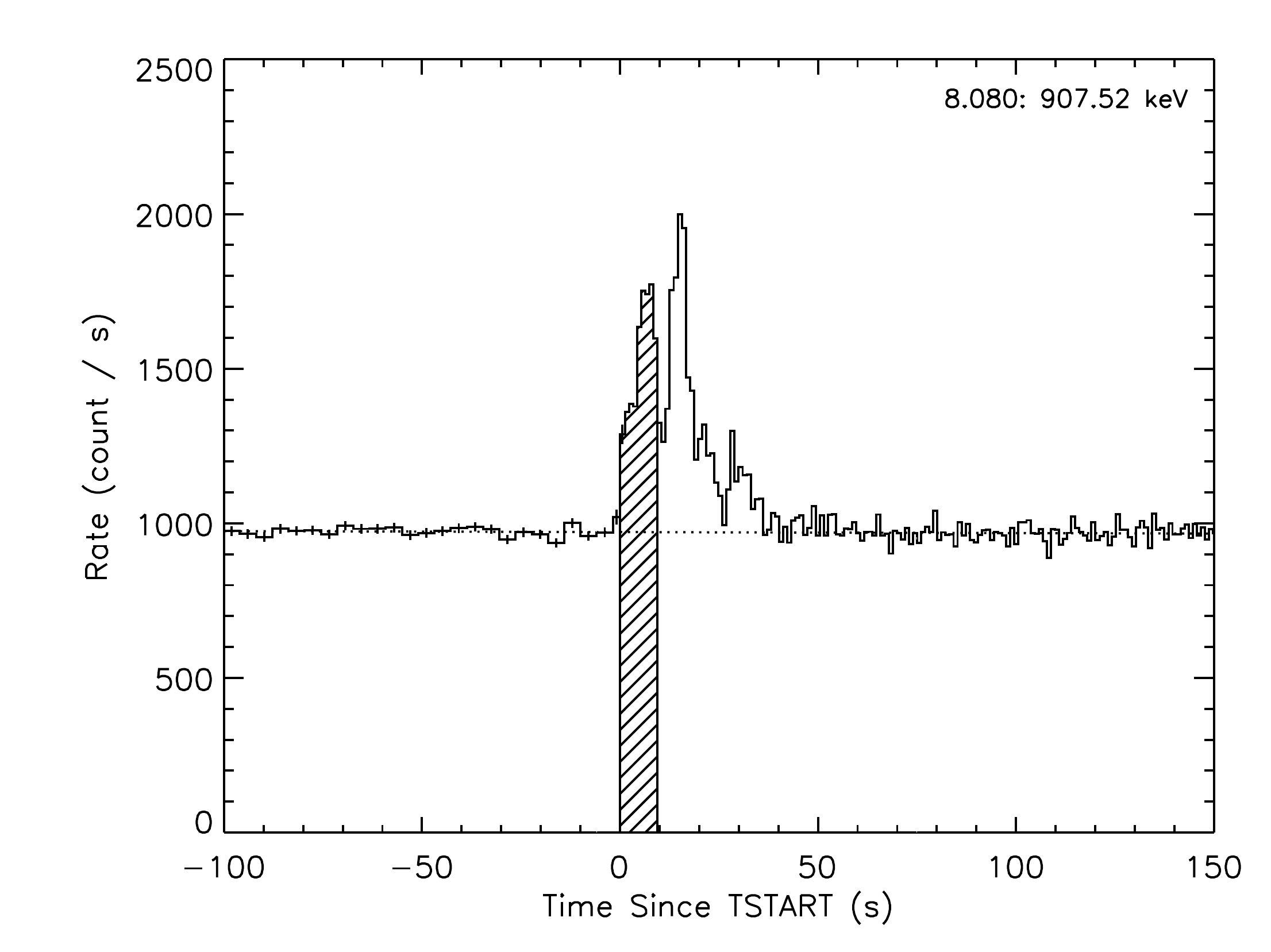}
        \end{subfigure}%
		\begin{subfigure}[]
                \centering
                \includegraphics[width=9cm]{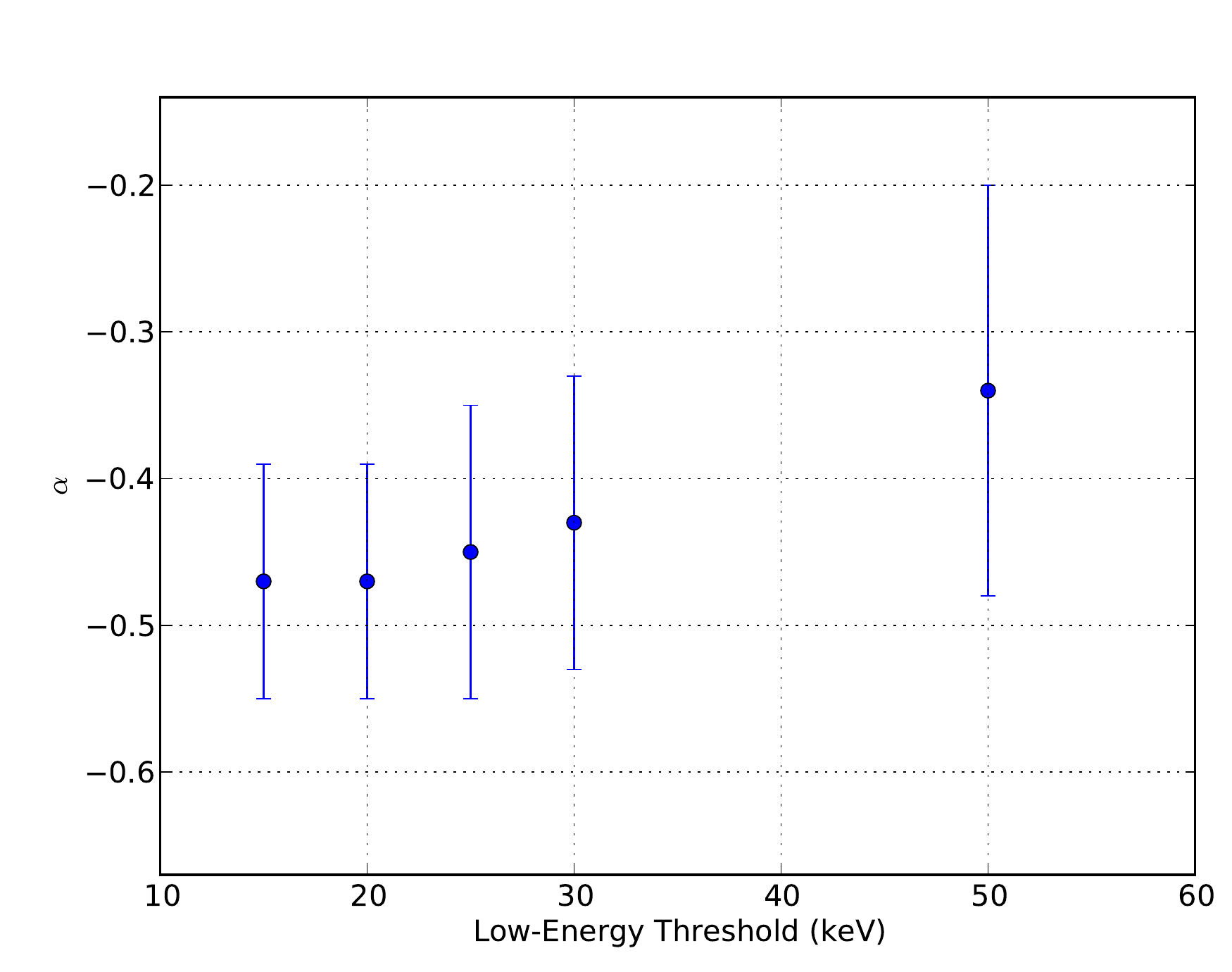}
        \end{subfigure}%
%\end{figure*}		
%
%\setcounter{figure}{5}		
%\noindent{}
%\begin{figure*}
\centering		
		\begin{subfigure}[]
                \centering
                \includegraphics[width = 6.8cm, angle=90]{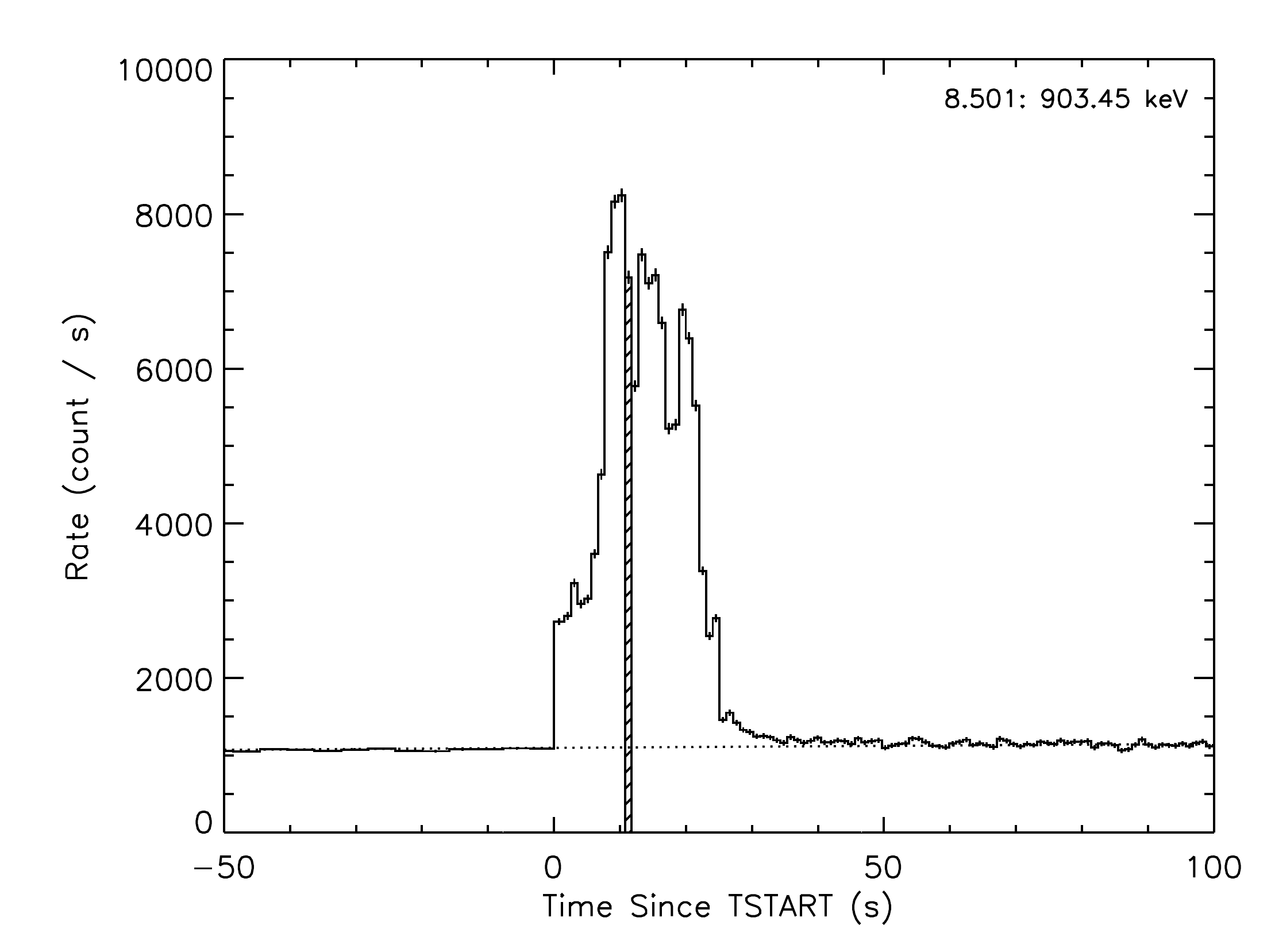}
        \end{subfigure}%
		\begin{subfigure}[]
                \centering
                \includegraphics[width=9cm]{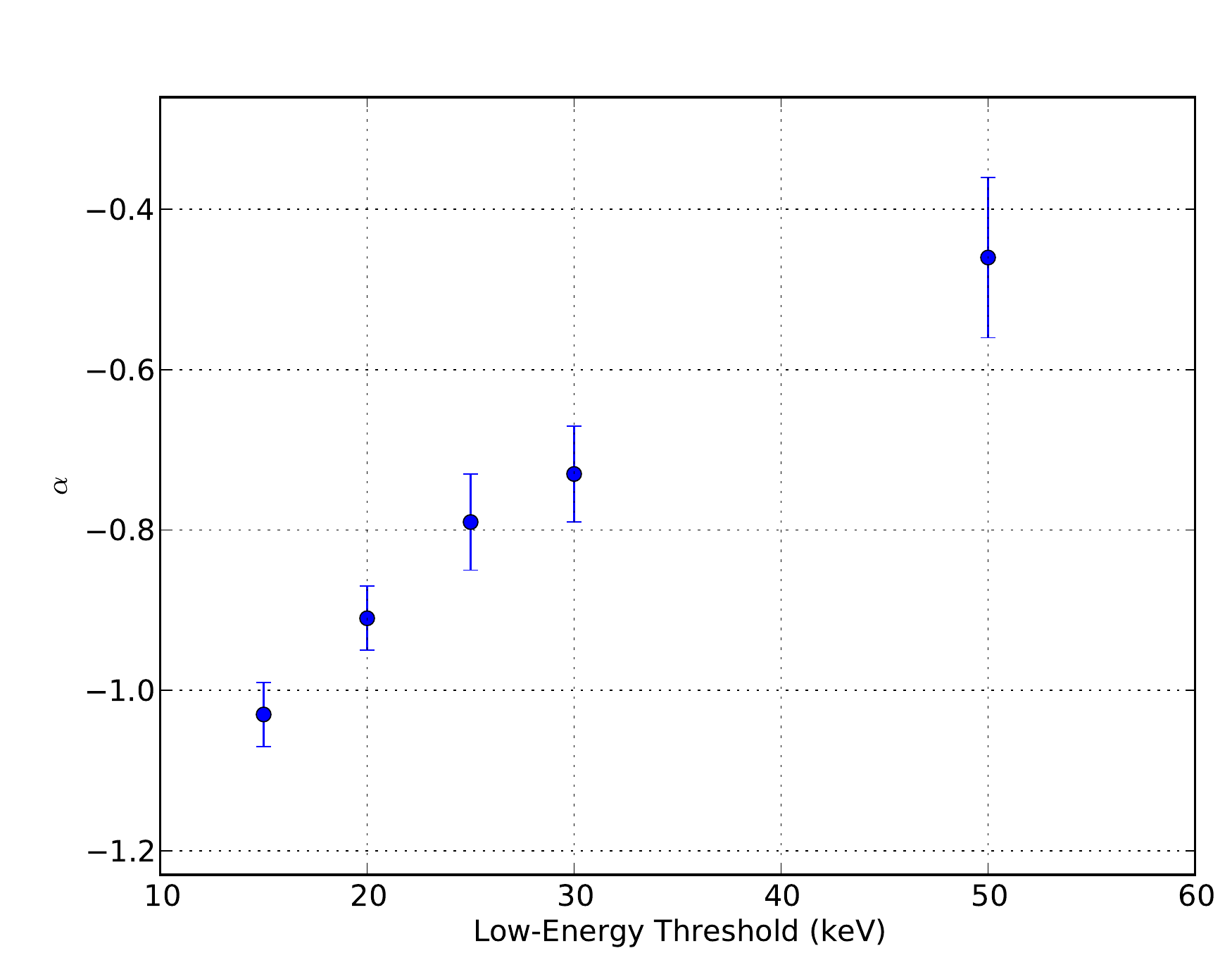}
        \end{subfigure}%
        \caption{Change in $\alpha$ between different LETs. The S/N temporal selections in GRB\,090102 and GRB\,090902B are presented in \textbf{a)} and \textbf{c)}. The spectral parameter $\alpha$ is presented with 2 $\sigma$ error bars in \textbf{b)} and \textbf{d)} for LET values from 15 keV to 50 keV. 
         $\alpha$ is consistent at the 2 $\sigma$ level as the LET is varied for the time interval selected in GRB\,090102 and is not consistent for the time interval in GRB\,090902B. 
      } 
\label{changeInalpha}
\end{figure*}

The second method involves looking at the variation in $\alpha$ while only changing the LET for the time-resolved analysis. The changes in $\alpha$ for one GRB which is consistent with a Band function and one which is not consistent are presented in Figure \ref{changeInalpha}. Two GRBs, GRB\,090618 and GRB\,091024 showed variations in $\alpha$ but did not have large deviations in the low-energy residuals. The parameter $\alpha$ varied outside 2 $\sigma$ during one temporal interval for each of the GRBs. These intervals were investigated further and were discounted due to a long, weak interval in GRB\,091024 and only having a single `good' detector available for analysis in GRB\,090618.

\noindent{}

Simulations were performed for GRBs that initially presented strong deviations in the time-integrated results. These deviations were then quantified by comparing the data to the simulated distribution for that GRB and were discarded as a chance occurrence if consistent with the simulations. Otherwise it was classified as a spectral deviation. A sample of GRBs were simulated that did not show strong deviations to ensure the veracity of the null hypothesis. 

Simulations could not be performed for time-resolved regions of all GRBs so the combination of the two methods outlined previously (summing the low-energy residuals and checking the change in $\alpha$ between different LETs) present the final criterion by which a deviation is tested. If a deviation passed both tests then it was classified as a real spectral anomaly. 

\subsection{Time-Integrated Spectra}
Out of the sample of 45 GRBs tested, a significant deviation in the time-integrated results was found in 3 cases GRB\,090902B (T0+0 - T0+25 s), GRB\,090424 (T0+0 - T0+6.4 s), GRB\,081215A (T0+0 - T0+8.4 s). A large 28.2 $\sigma$ excess (normalised) is noted in GRB\,090902B and is a clear outlier in Figure \ref{5SimGRBs}. GRB\,090424 has a deficit of 6.1 $\sigma$ in the time-integrated spectrum which appears to be due to an anomalous spectral feature near $E_{\rm peak}$. These two GRBs also show deviations in the time-resolved spectral analysis. GRB\,081215A manifests as a 6.7 $\sigma$ deficit but does not show up significantly in the time-resolved analysis. This deficit appears to be caused by strong spectral evolution throughout the GRB with $E_{\rm peak}$ decreasing from $>$ 1 MeV in the first 1 s time bin to $\sim$ 50 keV at the end of the 8 s emission region (a time-resolved analysis by \citet{2011ApJ...730..141Z} using different binning shows similar results). 

GRB\,090902B may be fit with an intense additional power-law component throughout the entire burst causing it to deviate strongly in the time-integrated spectrum \citep{2009ApJ...706L.138A, 2011ApJ...730..141Z}. GRB\,090424 has an intense additional component, which can be modelled as a blackbody throughout the burst which also manifests as a deviation in the time-integrated results. 
Additional emission processes and components may always be present in all GRBs but the instrument may not be sensitive enough to confidently detect them. 
A closer inspection of each GRB using time-resolved analysis is required to differentiate between artifacts in the spectra and real additional time-dependent spectral features.

\subsection{Time-Resolved Spectra}
A time-resolved analysis was performed on 672 spectra with an S/N of 25 $\sigma$ and 309 spectra with an S/N of 50 $\sigma$ which resulted in significant features in five bursts. GRB\,090323, GRB\,090424, GRB\,090820, GRB\,090902B and GRB\,090926A contained at least one temporal interval where significant deviations were detected.  Simulations were performed on the relevant time regions to ensure that they were significant. 
The properties of the five bursts are presented in Table \ref{FluenceProp}. 

\begin{table*}
\begin{minipage}{\textwidth}
\begin{center}
\caption{Properties of the GRBs found to have strong spectral deviations in the time-resolved analysis at low energies. } 
\begin{tabular}{llllrr}
\hline
GRB & Fluence\tablefootmark{a} & Fluence Rank\tablefootmark{a} & T90\tablefootmark{a} & Deviation\tablefootmark{b} & Localisation\tablefootmark{c}  \\
  &  10 - 1000 keV & first 2 years  &  50 - 300 keV  &  &   \\
  & (erg / cm$^{2}$) &  &  (s) & & \\
\hline
GRB\,090323 & $1.2\times10^{-4}\pm1.7\times10^{-7}$& 5 & 135.2$\pm$1.5 & +4.9 $\sigma$ (Excess)&  \textit{Swift} XRT\tablefootmark{1}  \\
GRB\,090424 & $4.6\times10^{-5}\pm3.9\times10^{-8}$& 19 & 14.1$\pm$0.3 & -6.2 $\sigma$ (Deficit)&  \textit{Swift} XRT\tablefootmark{2} \\
GRB\,090820 & $1.5\times10^{-4}\pm1.8\times10^{-7}$ & 3 & 12.4$\pm$0.2 & -5.5 $\sigma$ (Deficit)& GBM\tablefootmark{3} \\
GRB\,090902B & $2.2\times10^{-4}\pm3.2\times10^{-7}$ & 2 & 19.3$\pm$0.3 & +25.0 $\sigma$ (Excess)& \textit{Swift} XRT\tablefootmark{4} \\
GRB\,090926A & $1.5\times10^{-4}\pm3.4\times10^{-7}$ & 4 & 13.8$\pm$0.3 & +6.5 $\sigma$ (Excess)& \textit{Swift} XRT\tablefootmark{5} \\
\hline
\end{tabular}
\tablefoot{
(a) From \citet{2012ApJS..199...18P}.
(b) The deviations from the extrapolated fitting method have been normalised by the standard deviation from the simulated distributions for each individual burst section. 
(c) The \textit{Swift} XRT location is sufficiently accurate to model the response of the instrument for spectral analysis, even in cases where a more accurate location is known.
}
\tablebib{(1) \citet{2009GCN..9024....1K}; (2)  \citet{2009GCN..9223....1C}; (3)  \citet{2009GCN..9829....1C}; (4)  \citet{2009GCN..9868....1K}; (5)  \citet{2009GCN..9936....1V}
}
\label{FluenceProp}
\end{center}
\end{minipage}
\end{table*}

Out of the five bursts observed with low-energy deviations, four are in the top five most fluent GRBs from the first 2 years. In the case of the highest fluence event (GRB\,090618 with a fluence of $2.68\times10^{-4}\pm4.29\times10^{-7}$ erg / cm$^{2}$ \citep{2012ApJS..199...18P}), only one NaI was available for analysis due to the source-instrument geometry. It cannot be claimed with certainty that the result is stable on the basis of one `good' NaI detector. An analysis of this burst with the combined GBM-\textit{Swift} data did not significantly demonstrate the need for an additional component in the prompt gamma-ray emission \citep{2011MNRAS.416.2078P}.

As spectral evolution can lead to anomalous behaviour, even in the time-resolved analysis, lightcurves were examined for any notable features during a time interval which contained a deviation. To limit the effect of temporal binning, several further fits were performed by changing the length of the interval by up to $\pm$ 50 \% and by shifting the interval up to 50 \%. If the effects were still significant after this further analysis it is claimed as a detection of an additional spectral feature. These intervals were then fit with more complex models and the results are summarised in Table \ref{Spec_Table}. There is a notable improvement in the fit statistic when a Band function is fit with an additional component. This provides evidence that additional spectral features are needed to accurately model these time intervals. These solutions however are not unique and other spectral models may have a similar or better fit statistic.

\noindent{}
\begin{table*}
\caption{Time-resolved spectral parameters from model fits to GRB time intervals where significant deviations were detected.} 
\begin{tabular}{cccrrrrrrr}
\hline
GRB & Interval\tablefootmark{*} & Detectors & Model & Epeak & $\alpha$ & $\beta$ & kT & Index & C-Stat/DOF \\ %dt & Energy Flux (10 - 1000 keV) \\
\hline
\hline
GRB\,090323& 60.4:66.6 & n9+nb+b1 & Comp & 535.70$^{+33.00}_{-29.90}$ & -0.83$^{+0.03}_{-0.03}$ & - & - & - & 410.14/353 \\ %dt & (2.580 $\pm$ 0.055)$\times$10$^{-6}$ \\
&  &  & Band &532.00 $^{+35.30}_{-31.90}$ & -0.83$^{+0.03}_{-0.03}$ & -2.92$^{+0.37}_{-4.10}$ & - & - & 409.25/352 \\ %dt & (2.561 $\pm$  0.066)$\times$10$^{-6}$ \\
&  &  & Band+BB & 426.00 $^{+26.20}_{-23.50}$ & -0.52$^{+0.08}_{-0.07}$ & -3.00\tablefootmark{2} & 4.99$^{+0.53}_{-0.52}$ & - & 373.13/351 \\ %dt & (2.518 $\pm$ 0.059)$\times$10$^{-6}$ \\
&  &  & Band+PL & 409.20$^{+28.80}_{-26.30}$ & -0.28$^{+0.14}_{-0.13}$ & -3.00\tablefootmark{2} & - & -1.75$^{+0.08}_{-0.14}$ & 377.69/351 \\ %dt & (2.505 $\pm$  0.061)$\times$10$^{-6}$ \\
\hline
GRB\,090424& 2.3:3.3 & n6+n7+n8 & Comp & 169.50$^{+4.61}_{-4.40}$ & -0.87$^{+0.03}_{-0.03}$ & - & - & - & 738.26/596 \\ %dt & (6.325 $\pm$ 0.094)$\times$10$^{-6}$ \\
&  & +nb+b1 & Band & 153.00$^{+7.01}_{-7.33}$ & -0.80$^{+0.04}_{-0.04}$ & -2.81$^{+0.17}_{-0.23}$ & - & - & 729.21/595 \\ %dt & (6.557 $\pm$ 0.110)$\times$10$^{-6}$ \\
&  &  & Band+BB & 176.90$^{+10.90}_{-8.61}$ & -0.48$^{+0.12}_{-0.11}$ & -3.10$^{+0.23}_{-0.53}$ & 9.20$^{+0.55}_{-0.44}$ & - & 662.65/593 \\ %dt & (6.549 $\pm$ 0.110)$\times$10$^{-6}$ \\
&  &  & Band+PL & 153.40$^{+9.44}_{-9.44}$ & -0.80$^{+0.05}_{-0.00}$ & -2.82$^{+0.22}_{-0.22}$ & - & 0.16$^{+INF}_{-0.00}$ & 729.25/593 \\ %dt & (6.558 $\pm$ 0.120)$\times$10$^{-6}$ \\
\hline
GRB\,090820\tablefootmark{1}& 31.7:36.9 & n2+n5+b0 & Comp & 267.10$^{+2.36}_{-2.35}$ & -0.58$^{+0.01}_{-0.01}$ & - & - & - & 896.24/357 \\ %dt & (1.620 $\pm$ 0.009 )$\times$10$^{-5}$ \\
&  &  & Band & 221.60$^{+3.46}_{-3.45}$ & -0.45$^{+0.01}_{-0.01}$ & -2.63$^{+0.04}_{-0.04}$ & - & - & 667.45/356 \\ %dt & (1.622 $\pm$ 0.009)$\times$10$^{-5}$ \\
&  &  & Band+BB & 317.20$^{+9.24}_{-9.44}$ & -0.71$^{+0.02}_{-0.02}$ & -3.37$^{+0.18}_{-0.26}$ & 28.90$^{+0.91}_{-0.95}$ & - & 600.59/354 \\ %dt & (1.634 $\pm$ 0.009)$\times$10$^{-5}$ \\
&  &  & Band+PL & - & - & - & - & - & - \\ %dt & - \\
\hline
GRB\,090902B\tablefootmark{3}& 9.7:10.6 & n0+n1+b0 & Comp & 1695.00$^{+74.50}_{-85.30}$ & -1.14$^{+0.01}_{-0.01}$ & - & - & - & 1178.7/356 \\ %dt & ( 1.748 $\pm$ 0.020)$\times$10$^{-5}$ \\
&  &  & Band & 1657.00$^{+97.30}_{-66.80}$ & -1.14$^{+0.01}_{-0.00}$ & -3.00\tablefootmark{2} & - & - & 1198.0/356 \\ %dt & (1.739 $\pm$ 0.020)$\times$10$^{-5}$ \\
&  &  & Band+BB & 917.30$^{+37.70}_{-34.40}$ & -0.51$^{+0.04}_{-0.04}$ & -3.00\tablefootmark{2} & 5.01$^{+0.15}_{-0.16}$ & - & 489.48/354 \\ %dt & ( 2.047$\pm$ 0.027)$\times$10$^{-5}$ \\
&  &  & Band+PL & 782.00$^{+33.50}_{-27.70}$ & 0.20$^{+0.11}_{-0.12}$ & -3.00\tablefootmark{2} & - & -2.02$^{+0.04}_{-0.06}$ & 419.13/354 \\ %dt & (2.130 $\pm$ 0.029 )$\times$10$^{-5}$ \\
\hline
GRB\,090926A& 9.5:10.5 & n3+n6+n7 & Comp & 331.90$^{+10.10}_{-9.46}$ & -0.95$^{+0.02}_{-0.02}$ & - & - & - & 639.49/478 \\ %dt & (1.326 $\pm$ 0.018)$\times$10$^{-5}$ \\
&  & +b1 & Band & 311.40$^{+11.90}_{-11.50}$ & -0.93$^{+0.02}_{-0.02}$ & -2.64$^{+0.13}_{-0.19}$ & - & - & 628.82/477 \\ %dt & (1.314 $\pm$ 0.017)$\times$10$^{-5}$ \\
&  &  & Band+BB & 392.00$^{+44.50}_{-41.30}$ & -1.17$^{+0.05}_{-0.04}$ & -2.35$^{+0.11}_{-0.14}$ & 48.99$^{+3.54}_{-2.81}$ & - & 592.81/475 \\ %dt & (1.293 $\pm$ 0.018 )$\times$10$^{-5}$ \\
&  &  & Band+PL & 265.70$^{+11.70}_{-11.20}$ & -0.46$^{+0.09}_{-0.09}$ & -3.35$^{+0.41}_{-1.09}$ & - & -1.73$^{+0.04}_{-0.04}$ & 583.77/475 \\ %dt & (1.281 $\pm$0.017 )$\times$10$^{-5}$ \\
\hline
\end{tabular}
\tablefoot{
*)  Measured in seconds since trigger time (T0).
1)  The Band+PL fit for GRB\,090820 did not converge to physical parameters and is omitted. 
2)  $\beta$ was frozen to -3.00 if the uncertainties were unconstrained.
3)  Although many intervals have significant excesses in GRB\,090902B, only one is presented above for illustrative purposes \citep[See also][]{2009ApJ...706L.138A,2012MNRAS.420..468P,2011MNRAS.415.3693R}.
}
\label{Spec_Table}
\end{table*}

\subsection{Individual Bursts with anomalous low-energy spectra }
Significant anomalous spectral behaviour with respect to a Band function was found in time intervals of five bursts. The spectra were further investigated by fitting a number of spectral models as presented in Table\,\ref{Spec_Table}. The individual events are described in more detail below.

\subsubsection{GRB\,090323}
\begin{figure*}
\centering
		\begin{subfigure}[]
                \centering
                \includegraphics[width = 6.8cm, angle=90]{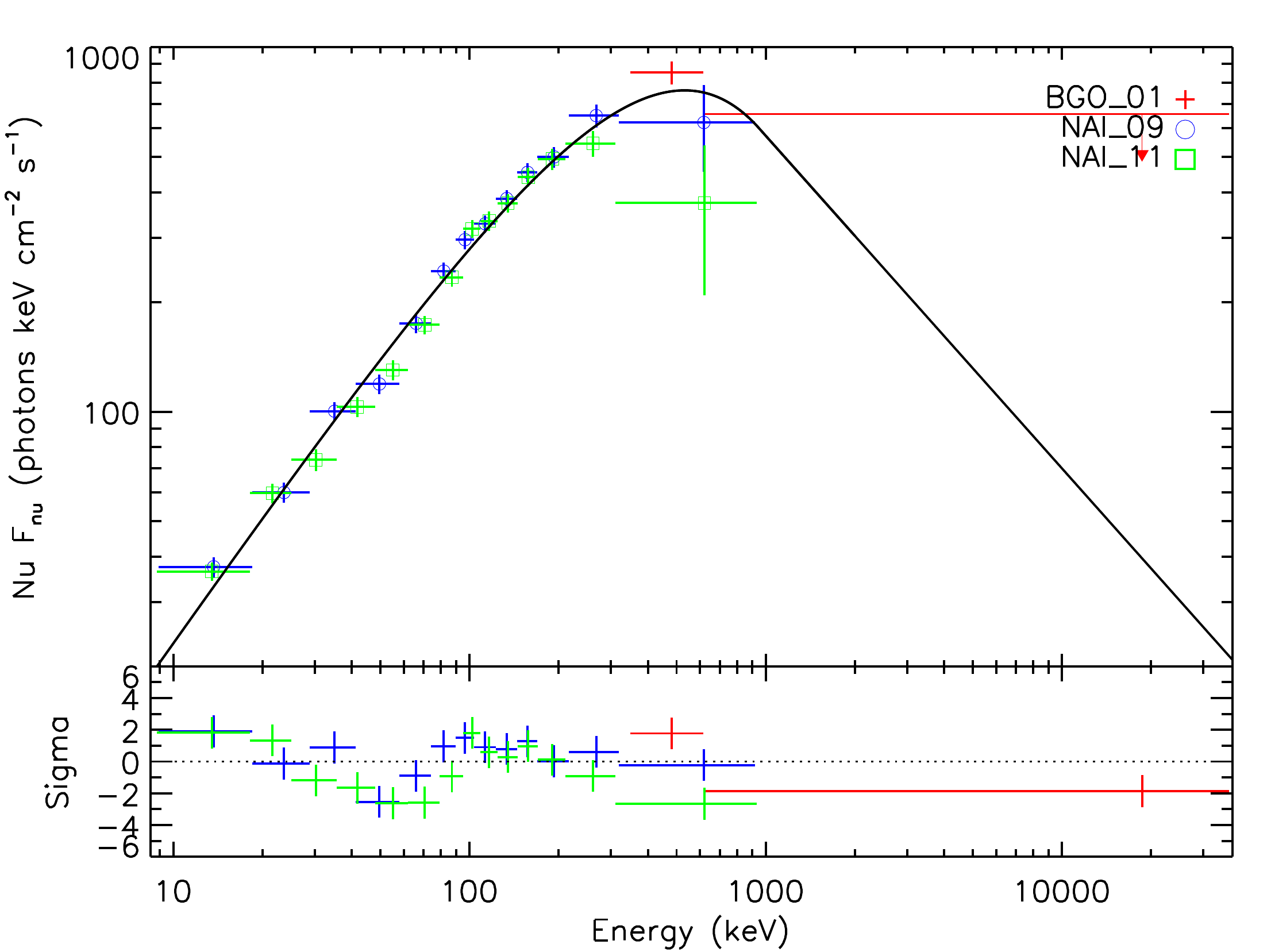}
        \end{subfigure}%
		\begin{subfigure}[]
                \centering
                \includegraphics[width = 6.8cm, angle=90]{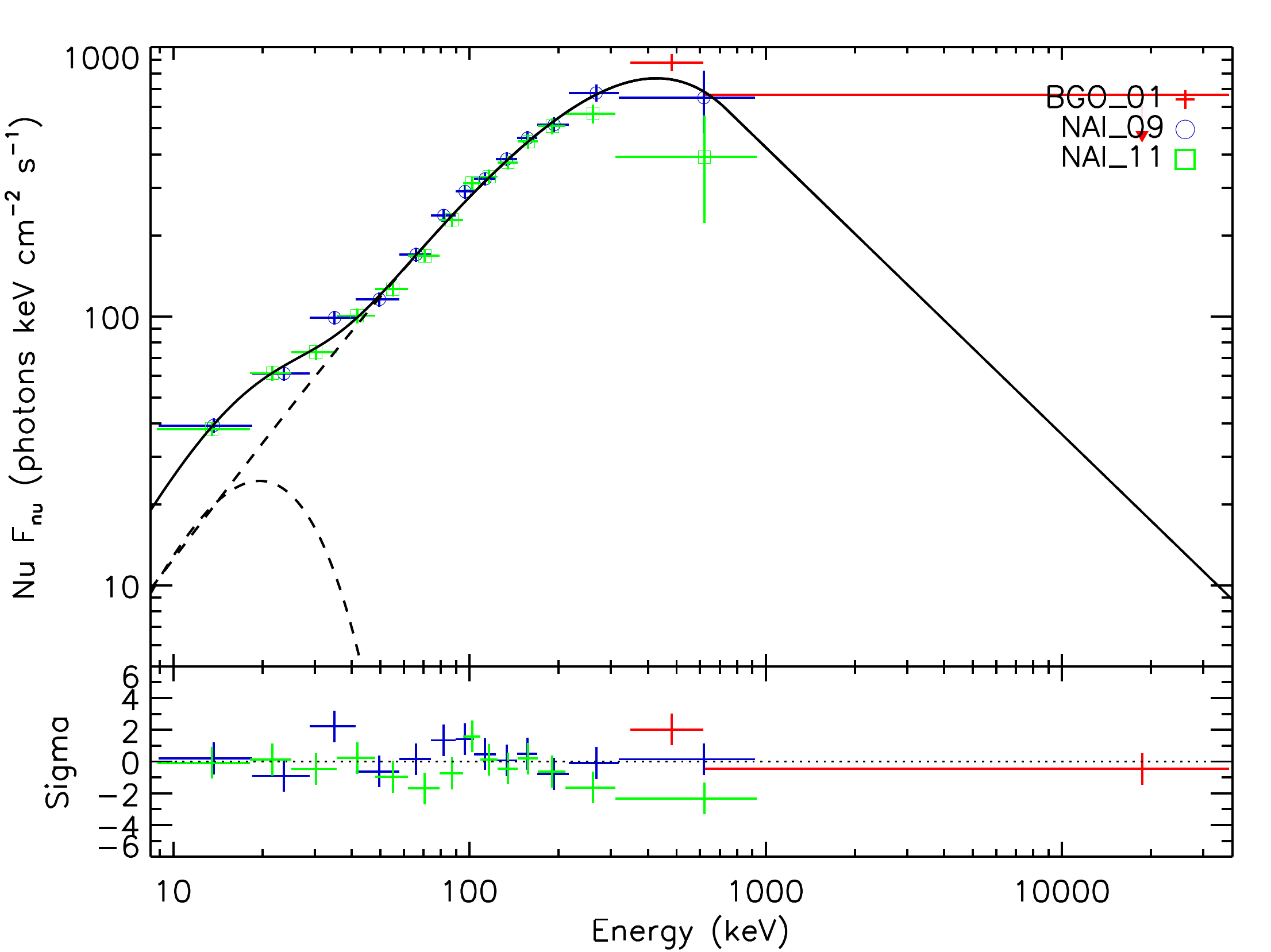}
        \end{subfigure}%
		\begin{subfigure}[]
                \centering
                \includegraphics[width = 6.8cm, angle=90]{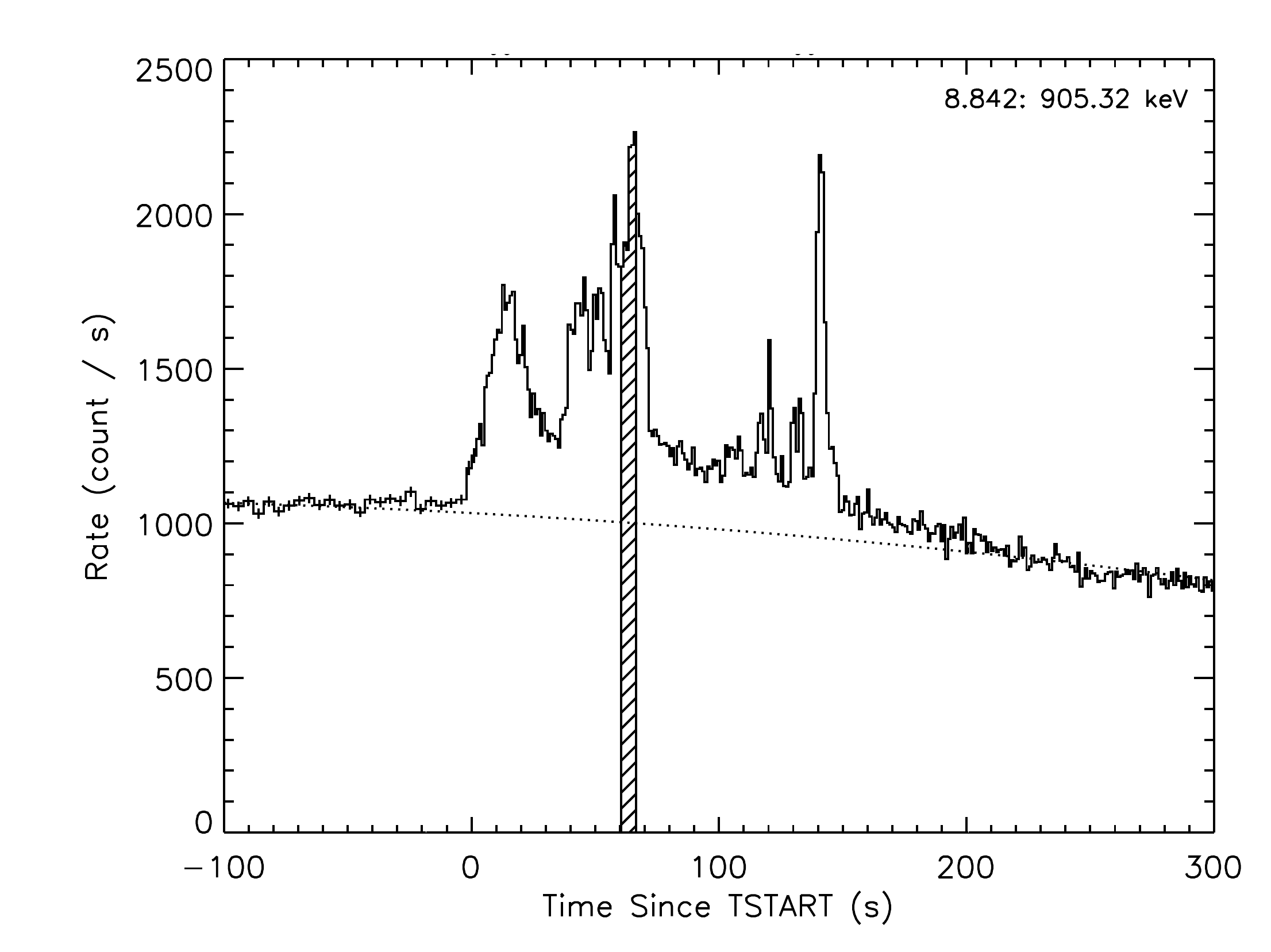}
        \end{subfigure}%
\caption{Spectral fits for GRB\,090323. \textbf{a)} A Band spectral fit and \textbf{b)} a Band + blackbody spectral fit are displayed for the hatched region of interest in \textbf{c)} the CSPEC lightcurve. All fits were performed on the full resolution data and the data were binned for illustrative purposes. The fit parameters are given in Table \ref{Spec_Table}.}
\label{090323}
\end{figure*}

GRB\,090323 was detected by \textit{Fermi} GBM \citep{2009GCN..9035....1V} and LAT which caused the satellite to perform an autonomous repoint to the source location \citep{2009GCN..9021....1O}. The lightcurve from GBM NaI detector n9 is displayed in Fig.\,\ref{090323} and consists of multiple distinct pulses over $\sim$ 150 s. Multi-wavelength observations were aided by an on-ground location provided by the LAT and X-ray observations from Swift-XRT \citep{2009GCN..9024....1K}. Follow-up observations report a spectroscopic redshift of  z = 3.57 \citep{2009GCN..9028....1C} and the afterglow was also detected at radio wavelengths \citep{2009GCN..9043....1H}. The energetics and host galaxy properties of this burst are described in \citet{2010A&A...516A..71M}.

After performing a spectral analysis on this burst an excess is detected in two out of nine intervals (52 - 60 s, 60 - 66 s relative the trigger time). The spectral analysis for the interval with the strongest low-energy deviation is presented in Figure \ref{090323} and Table \ref{Spec_Table}. An improvement in the spectral residuals is notable when a Band + blackbody or Band + power-law fit is performed over a single Band fit. 

\noindent{}

\subsubsection{GRB\,090424}

The prompt emission from GRB\,090424 was detected by Swift-BAT, XRT and UVOT which provided an arcsecond localisation to the follow-up community within minutes \citep{2009GCN..9223....1C}. The prompt emission was also observed by \textit{Fermi} GBM \citep{2009GCN..9230....1C} and Suzaku WAM \citep{2009GCN..9270....1H}. Multiple optical telescopes observed the optical afterglow (e.g. \citep{2009GCN..9245....1O}) and a spectroscopic redshift of z = 0.544 was obtained by Gemini-South \citep{2009GCN..9243....1C}. Radio observations detected a bright radio afterglow \citep{2009GCN..9260....1C} and a disturbance in Earth's ionosphere was detected by monitoring very low frequency radio waves \citep{2011GCN..11883...1M, 2010InJPh..84.1461C}.

\begin{figure*}
\centering
		\begin{subfigure}[]
                \centering
                \includegraphics[width = 6.8cm, angle=90]{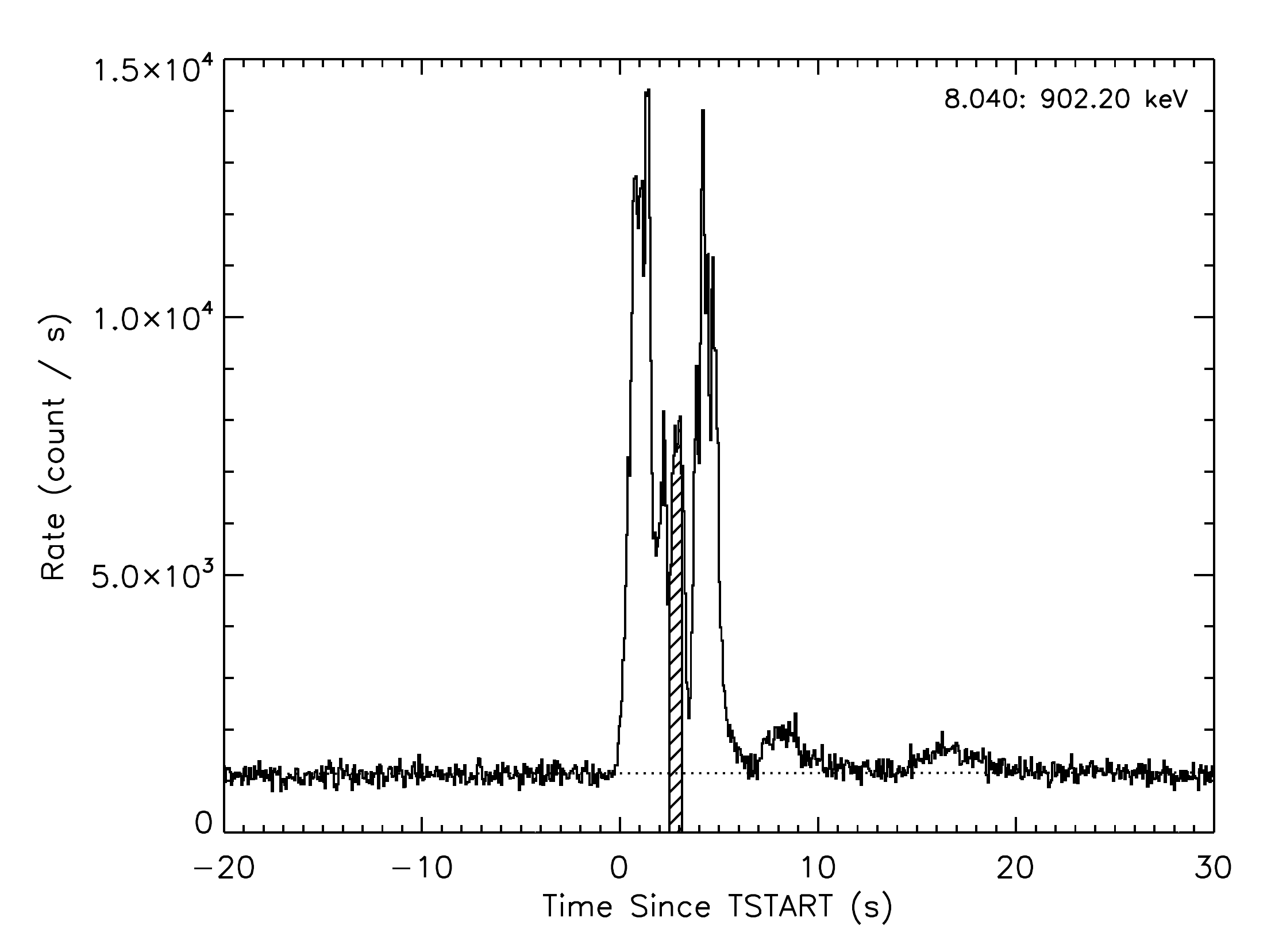}
        \end{subfigure}%
		\begin{subfigure}[]
                \centering
                \includegraphics[width = 6.8cm, angle=90]{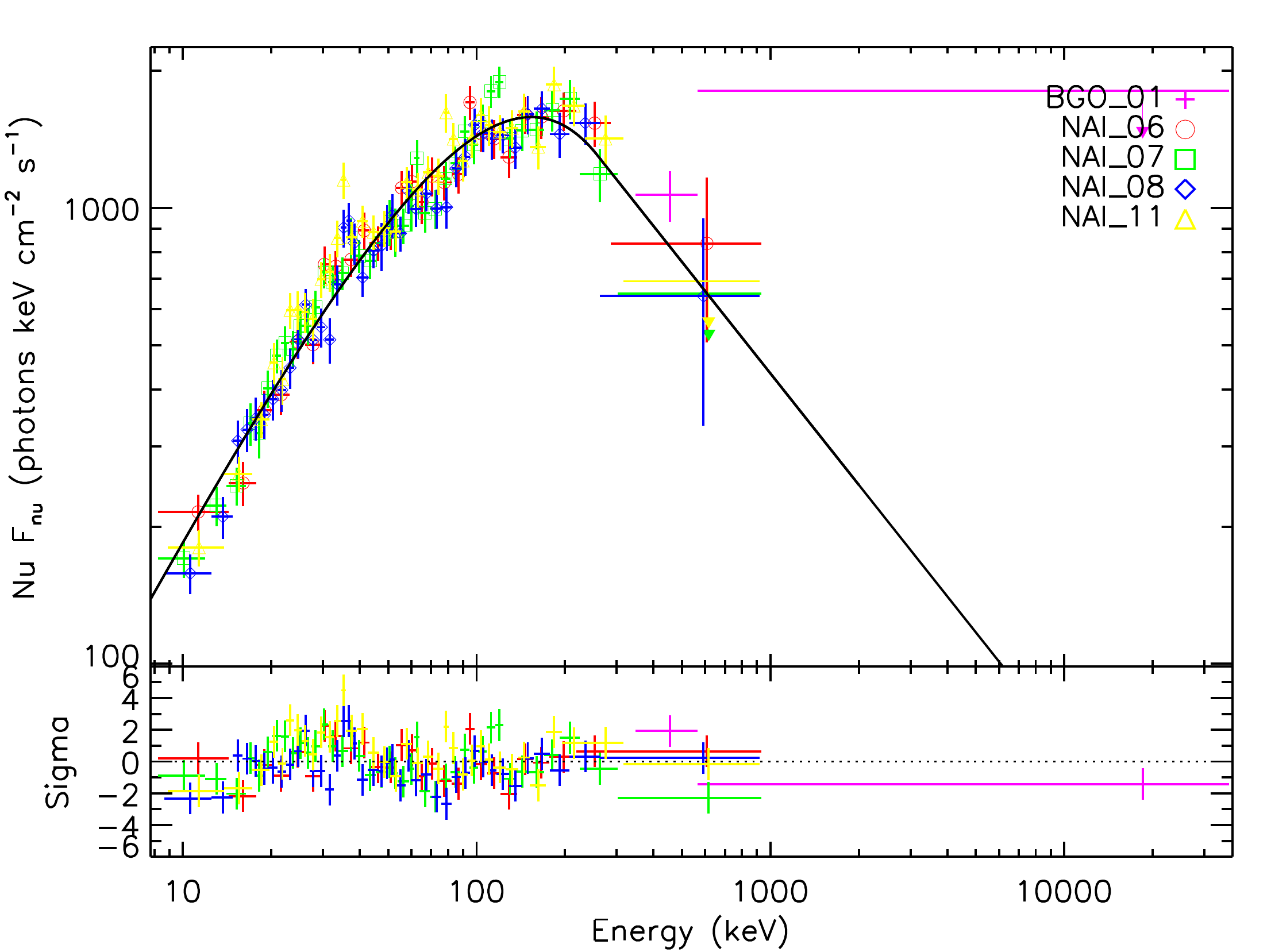}
        \end{subfigure}%
%        \end{figure*}
%\setcounter{figure}{7}      
%\noindent{}
%\begin{figure*}
\centering        
		\begin{subfigure}[]
                \centering
                \includegraphics[width = 6.8cm, angle=90]{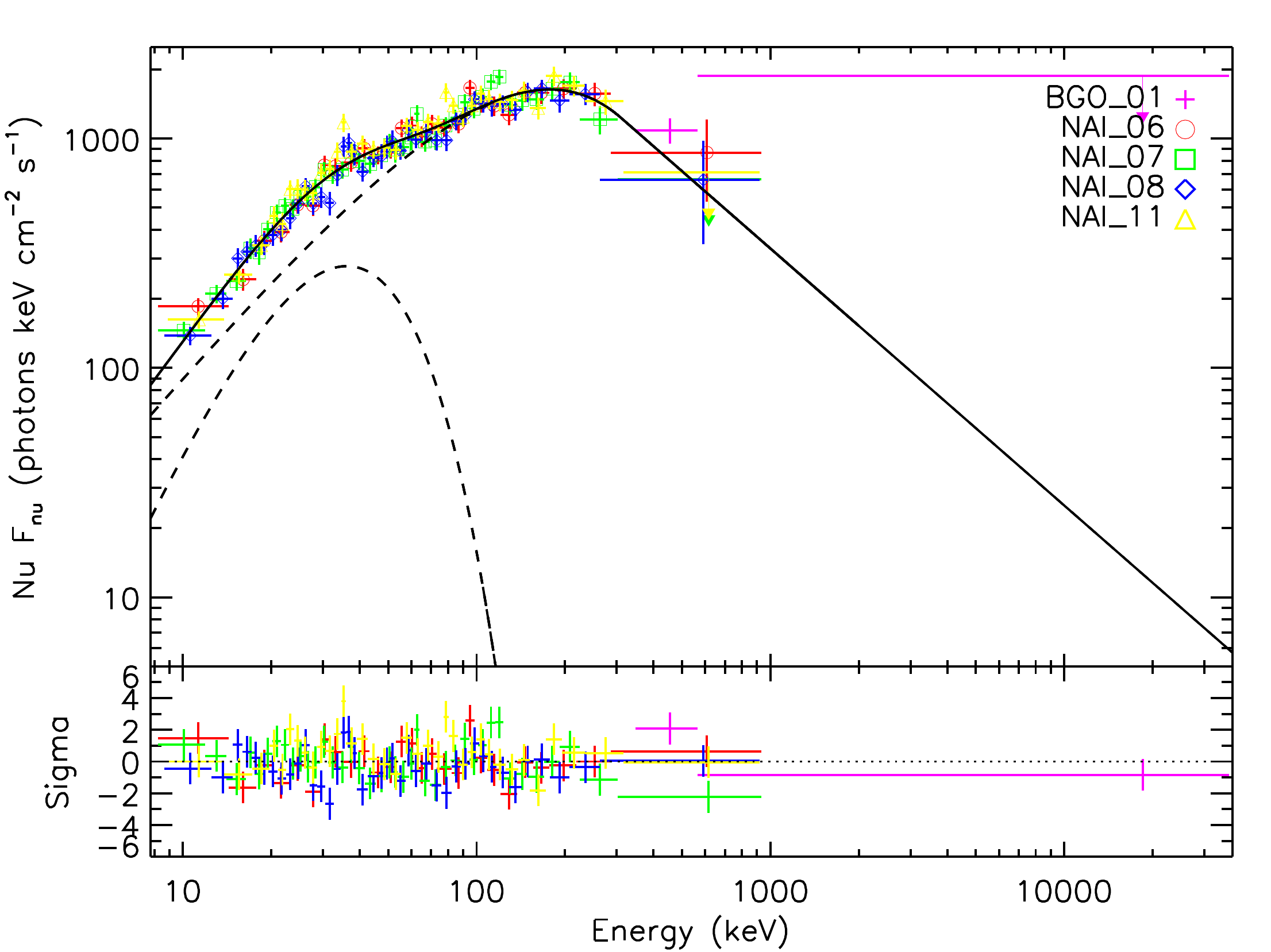}
        \end{subfigure}%
		\begin{subfigure}[]
                \centering
                \includegraphics[width = 6.8cm, angle=90]{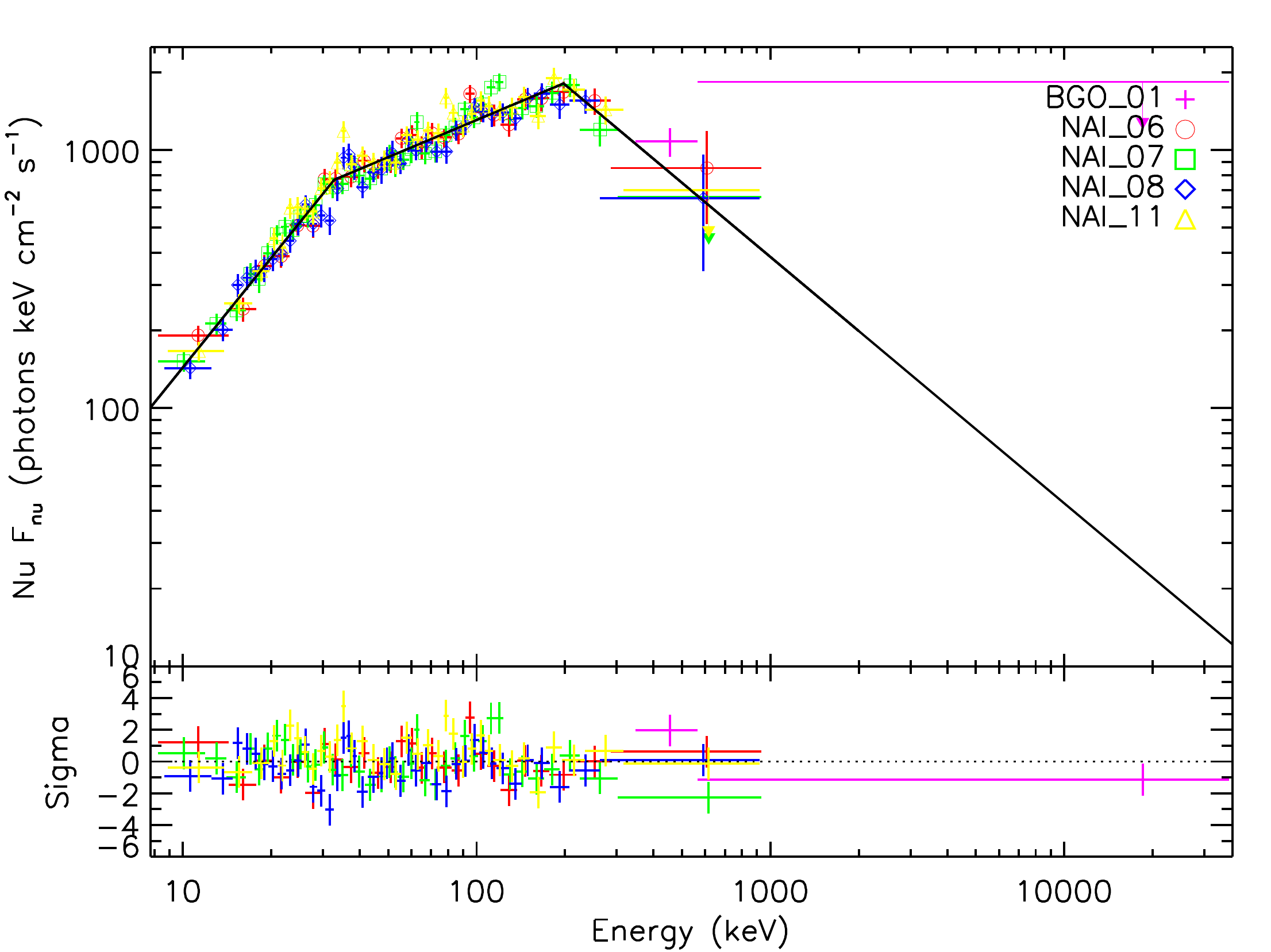}
        \end{subfigure}%
\caption{Spectal fits for GRB\,090424. \textbf{a)} The high-time resolution lightcurve of GRB\,090424 with the hatched region being the region of interest. \textbf{b)} A simple Band fit to selected region. \textbf{c)} A Band + blackbody fit to the data. \textbf{d)} A double broken power-law fit with parameters of E$_{\rm break1}$ = 32.74$^{+1.56}_{-1.55}$ keV, E$_{\rm break2}$ = 197.70$^{+14.10}_{-15.10}$ keV, Index1 = -0.59$^{+0.07}_{-0.06}$, Index2 = -1.52$^{+0.03}_{-0.03}$, Index3 = -2.95$^{+0.16}_{-0.16}$ (where Index1 $<$ E$_{\rm break1}$, E$_{\rm break1}$ $<$ Index2 $<$ E$_{\rm break2}$, Index3 $>$ E$_{\rm break2}$).}
\label{090424}
\end{figure*}

The main emission period of this GRB is relatively short compared to other bursts in the high fluence sample and has an extremely high peak flux of 110 photons~/~s \citep{2012ApJS..199...18P}. The burst has at least 3 pulses in rapid succession lasting $\sim$ 6 s and lower level emission that continues up to 20 s after the trigger time. A significant low-energy deficit was observed in the time-integrated and time-resolved spectral analysis of GRB\,090424. This apparent deficit is caused by the unusual spectral shape, which is shown in Figure \ref{090424}. The spectrum has 2 distinct spectral breaks, one in the usual energy range where a break due to $E_{\rm peak}$ is to be expected but also another lower energy break in the spectrum. A Band + blackbody fit with a blackbody kT of $\sim$ 9 keV can be fit throughout the burst which is most prominent between 2 - 3~s relative to the trigger time. Six 1~s intervals between 0 - 6~s were analysed and a strong deviation is present in the residuals in the the 2 - 3~s region. Little spectral evolution is present throughout the burst, and as this feature is observable throughout the burst it is less likely to be caused by spectral evolution.

\noindent{}

\subsubsection{GRB\,090820}

GBM detected GRB\,090820 and issued an automated repoint request to \textit{Fermi}, however the LAT could not observe the burst due to Earth avoidance constraints \citep{2009GCN..9829....1C}. The RT-2 Experiment onboard the CORONAS-PHOTON satellite also observed the prompt emission \citep{2009GCN..9833....1C}. No further follow-up of the burst occurred and thus the best location for the burst is obtained from GBM. The burst triggered on a weak precursor which was excluded from the spectral analysis. 

\begin{figure*}
\centering
		\begin{subfigure}[]
                \centering
                \includegraphics[width = 6.1cm, angle=90]{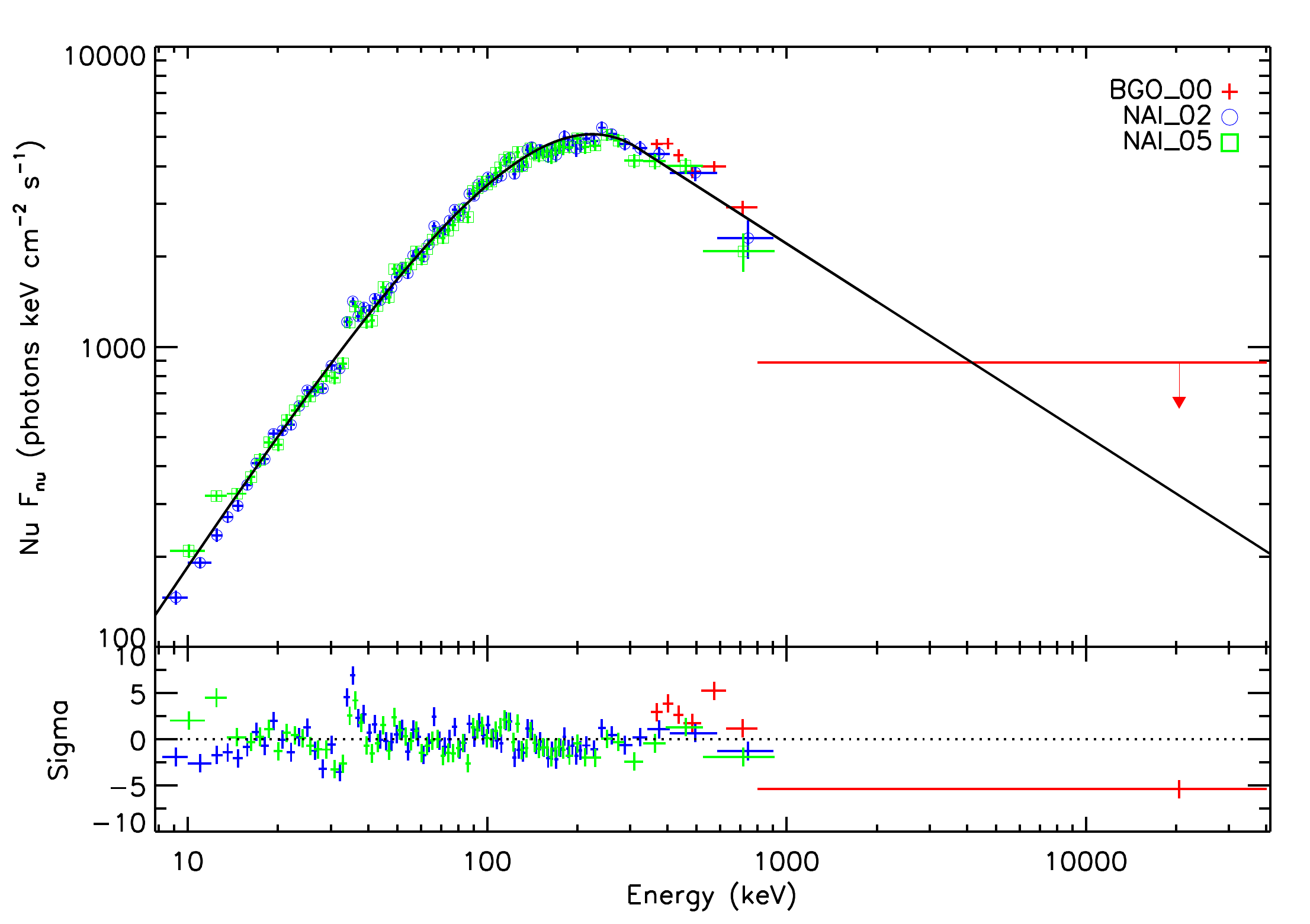}
        \end{subfigure}%
		\begin{subfigure}[]
                \centering
                \includegraphics[width = 6.1cm, angle=90]{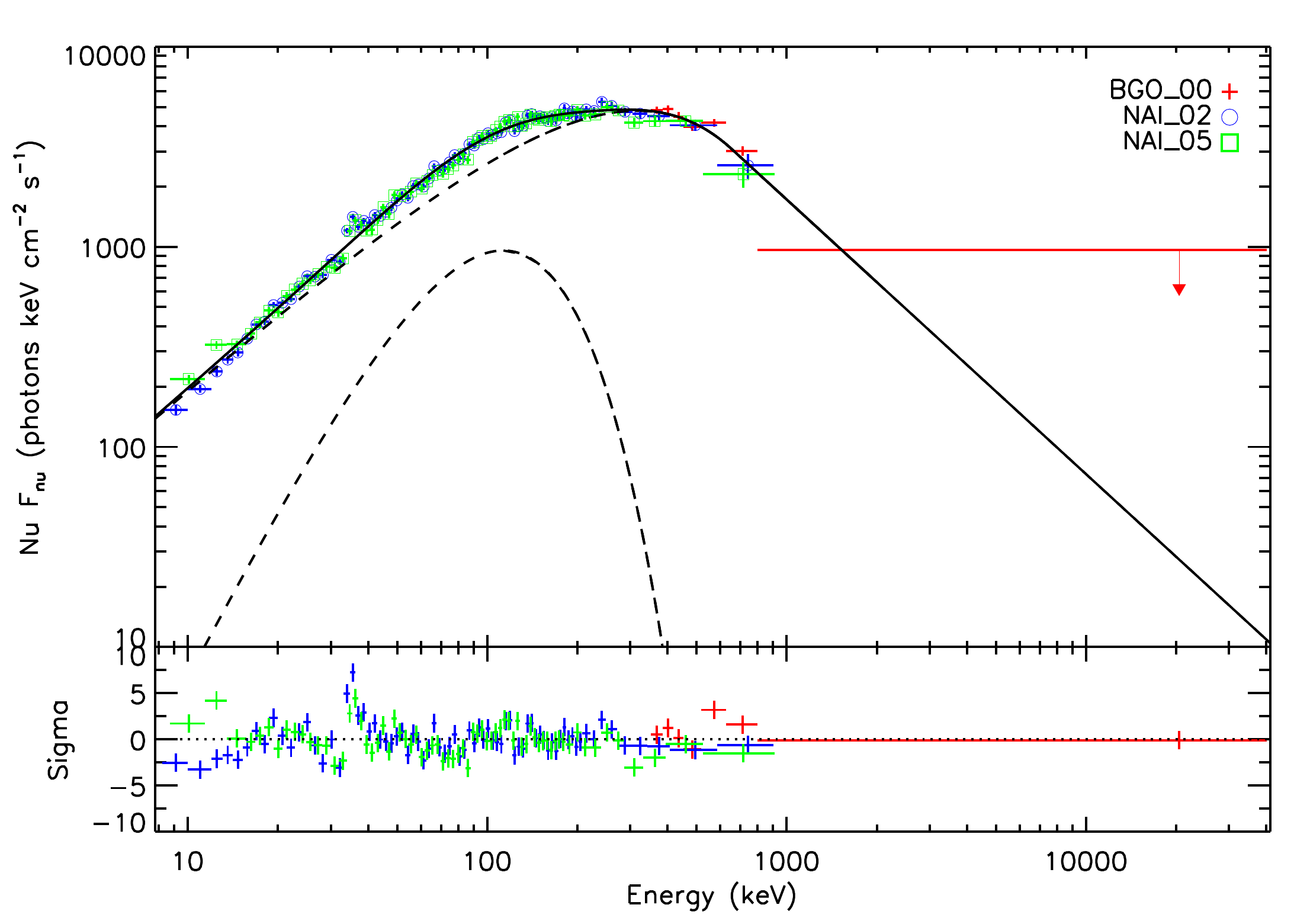}
        \end{subfigure}%
		\begin{subfigure}[]
                \centering
                \includegraphics[width = 6.1cm, angle=90]{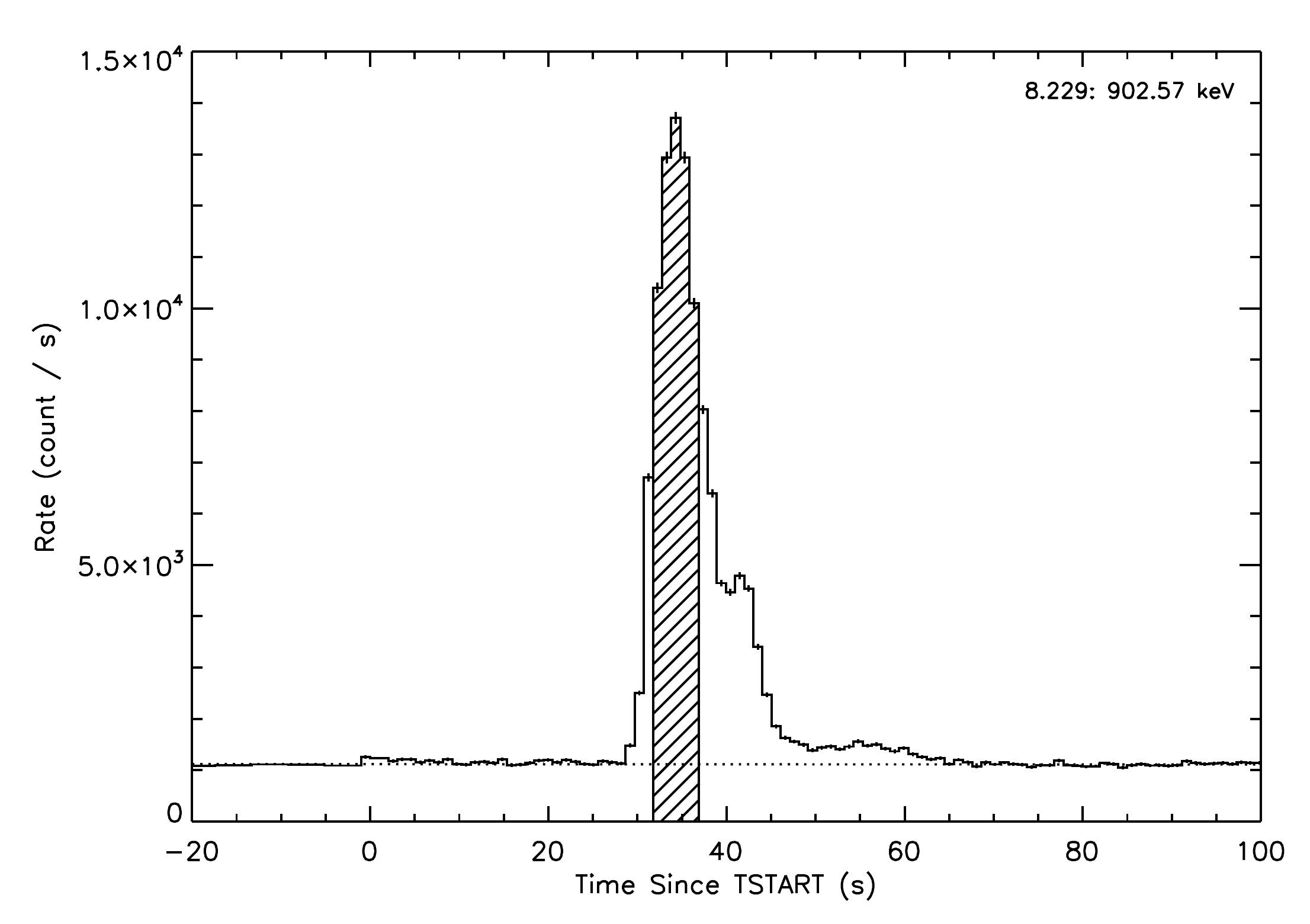}
        \end{subfigure}%
\caption{Spectral data for GRB\,090820. The data were fit with \textbf{a)} a Band function and \textbf{b)} a Band + blackbody function. \textbf{c)} The CSPEC lightcurve of GRB\,090820 with the hatched region being the region of interest.}
\label{090820}
\end{figure*}

This burst presents a low-energy deficit relative to a Band function which is caused by the Band function attempting to fit data that has a broader peak relative to the expected function fit. Thirteen time intervals between 29 - 45~s from the trigger time were analysed with five 1~s intervals between 31 - 36 s exhibiting marginal evidence of deviations. As this was the only GRB with consecutive marginal evidence for deviations, these time intervals were analysed together and display a strong deviation in the spectral data. By adding a blackbody component the Band function can account for the broader spectral peak as shown in Figure \ref{090820}. As an additional power-law could not account for deviations at lower energies, this provides evidence that some additional spectral property is occurring in the mid-energy range. Spectral analysis comparing standard models to more advanced fitting techniques, such as synchrotron models have been carried out by \citet{2011ApJ...741...24B} for this burst.

\noindent{}

\subsubsection{GRB\,090902B}

GRB\,090902B \citep{2009GCN..9866....1B} was an extremely bright GRB with a bright additional component observed across 8 orders of magnitude in energy \citep{2009ApJ...706L.138A}. The GRB was initially detected by both instruments on \textit{Fermi}, GBM and the LAT \citep{2009GCN..9866....1B,2009GCN..9867....1D}. The prompt emission was also detected by Suzaku WAM \citep{2009GCN..9897....1T}. A Target of Opportunity (ToO) was issued to \textit{Swift} which observed an uncataloged source within the LAT error radius \citep{2009GCN..9868....1K}. Numerous optical follow-up observations were made \citep[e.g.][]{2009GCN..9874....1O} resulting in a redshift of z = 1.822 \citep{2009GCN..9873....1C}. The source was also observed in the radio \citep{2009GCN..9883....1V}.

\begin{figure*}
\centering
		\begin{subfigure}[]
                \centering
                \includegraphics[width = 6.1cm, angle=90]{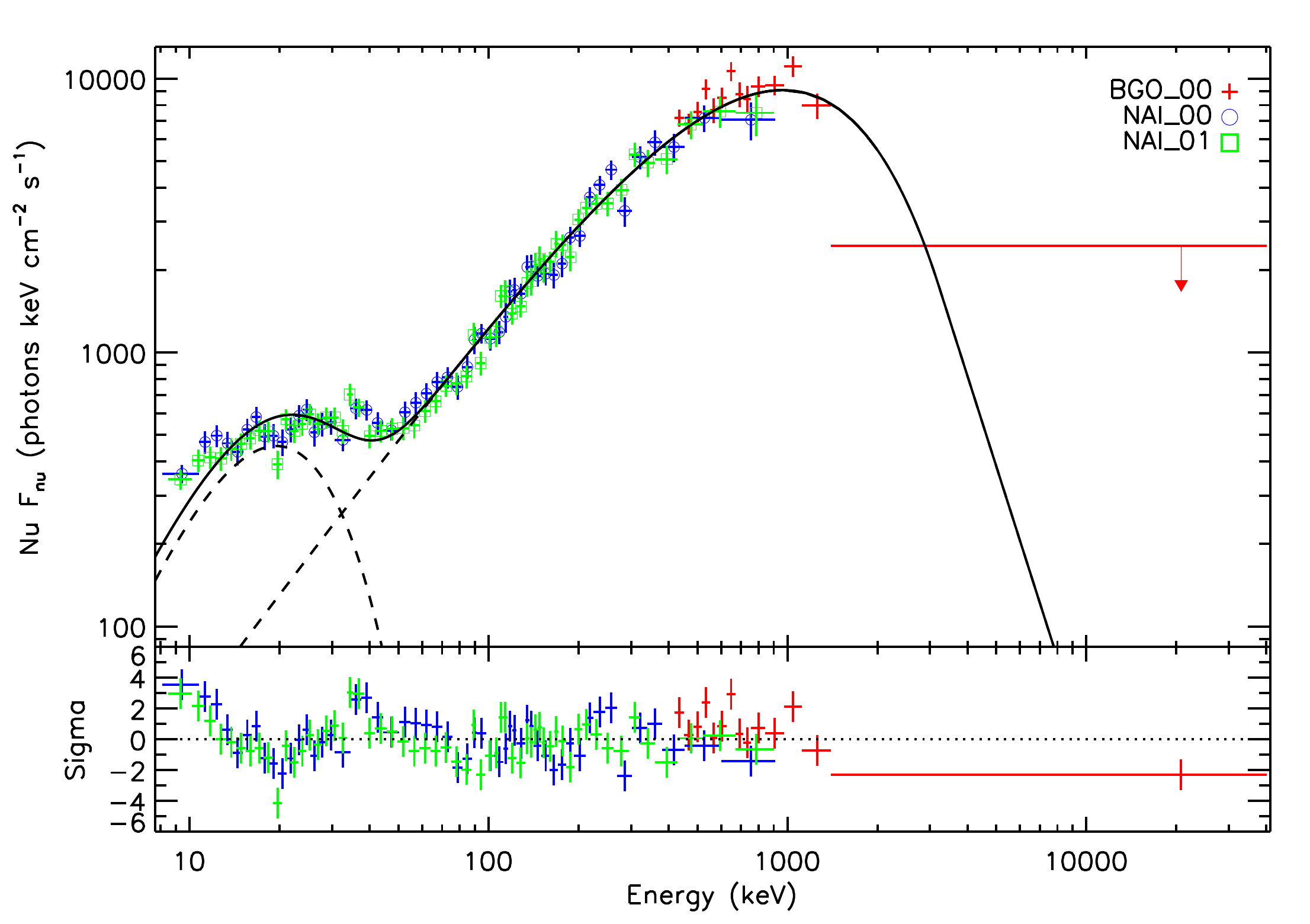}
        \end{subfigure}%
		\begin{subfigure}[]
                \centering
                \includegraphics[width = 6.1cm, angle=90]{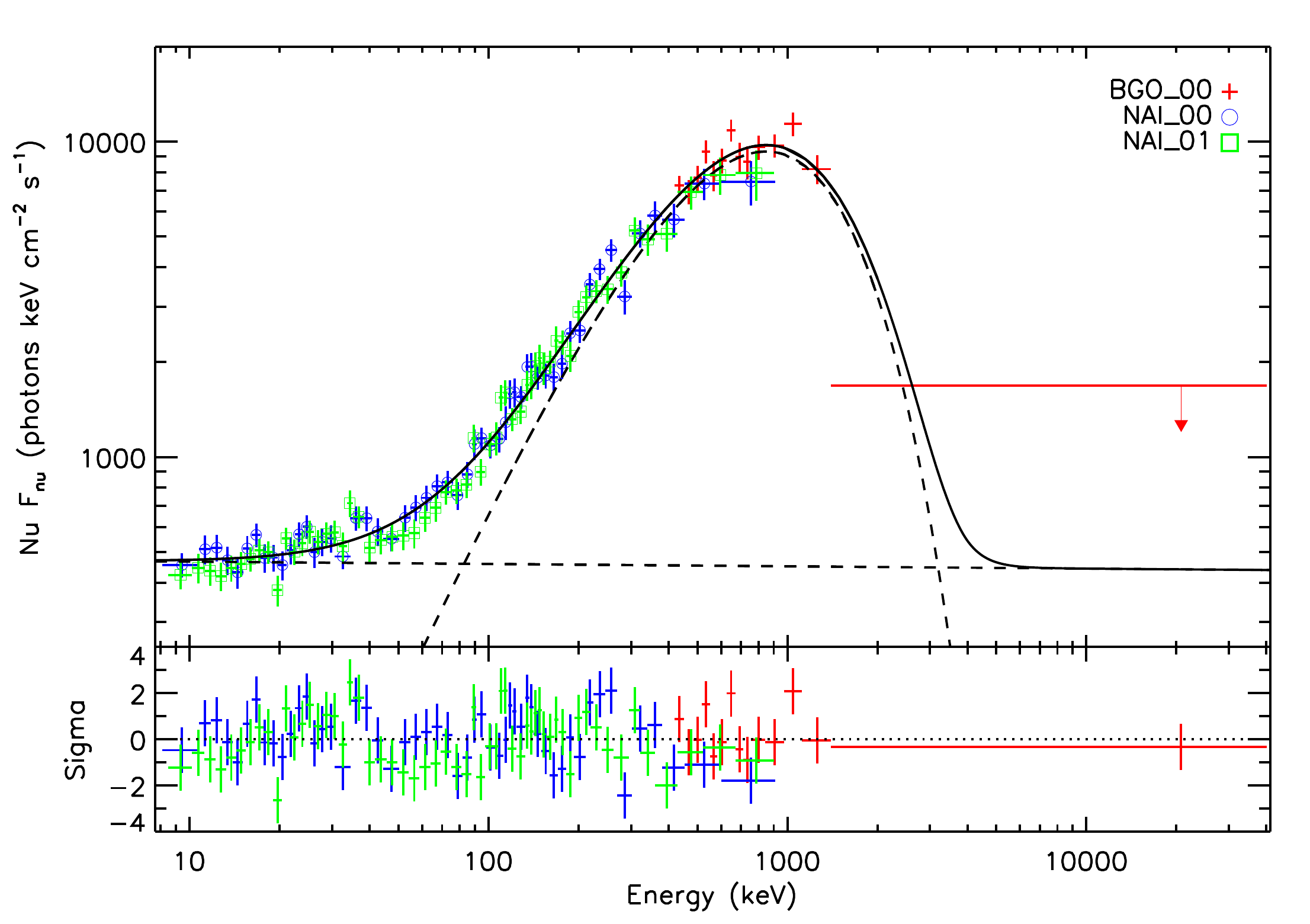}
        \end{subfigure}%
		\begin{subfigure}[]
                \centering
                \includegraphics[width = 6.1cm, angle=90]{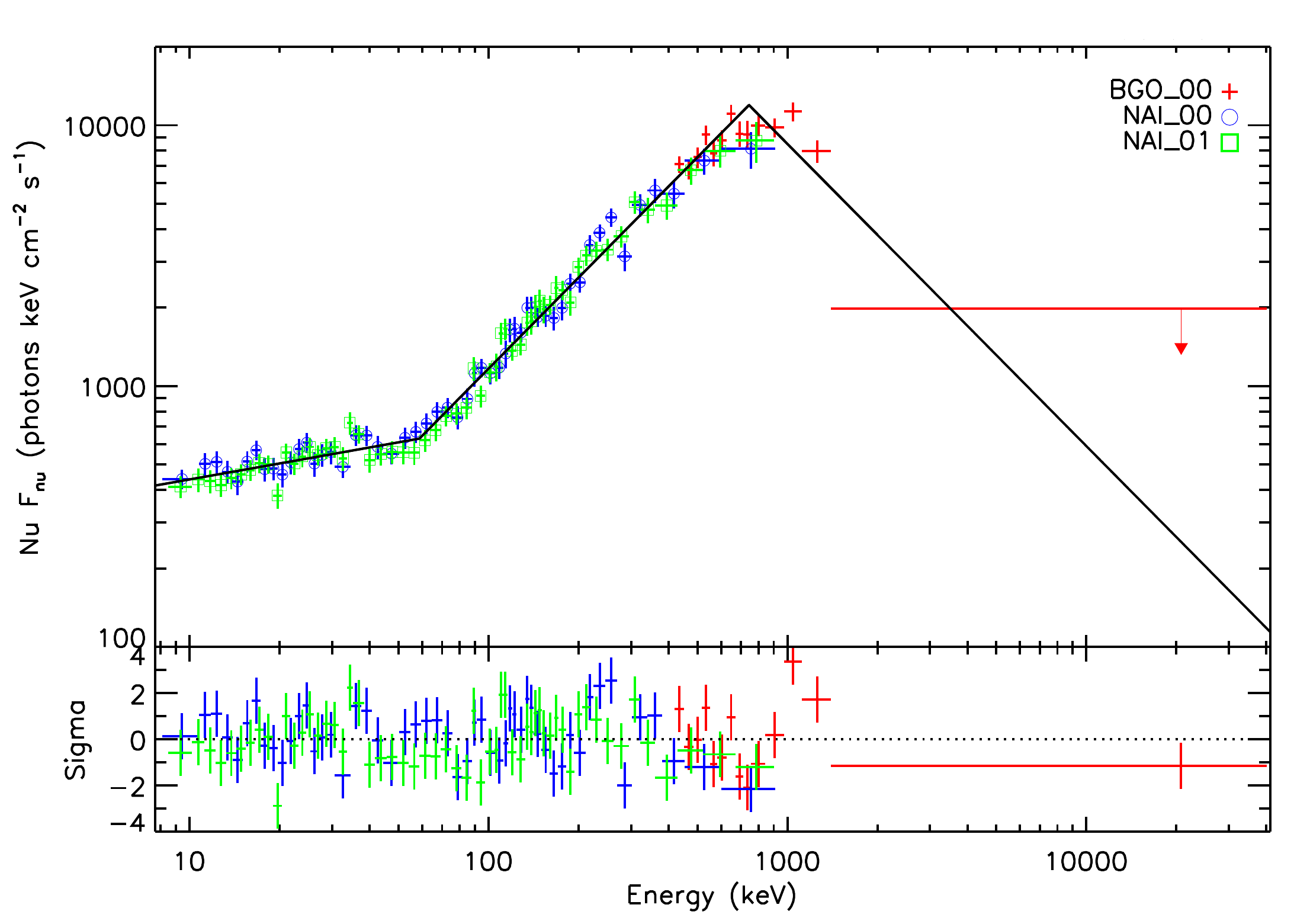}
        \end{subfigure}%
\caption{Spectral fits applied to the the time interval of T0+9.7 - T0+10.6 s for GRB\,090902B. Fits performed to decribe the data include; \textbf{a)} a Band + blackbody fit, \textbf{b)} Band + power-law fit and \textbf{c)} a double broken power-law fit with parameters of E$_{\rm break1}$ = 58.79$^{+3.25}_{-3.25}$ keV, E$_{\rm break2}$ = 743.60$^{+36.70}_{-39.90}$ keV, Index1 = -1.80$^{+0.03}_{-0.04}$, Index2 = -0.84$^{+0.03}_{-0.02}$, Index3 = -3.16$^{+0.14}_{-0.13}$ (see Figure \ref{090424} for description of parameters). A lightcurve is shown in Figure \ref{changeInalpha} (c).}
\label{090902BB}
\end{figure*}

A significant excess was observed in this GRB in numerous intervals (ten 1-second intervals between 6 - 16 s relative the trigger time). The region with the stongest deviation is presented in Table \ref{Spec_Table}. An additional power-law component was the best fit model to the data out of the models tested. The additional power-law index from analysing the GBM data only in this region, $\alpha$ = -2.00$^{+0.04}_{-0.06}$, is consistent with the power-law index observed by a joint fit over the entire spectral range of GBM and LAT ($\alpha$ = -1.98$^{+0.02}_{-0.02}$). Complex spectral models to this GRB are fit in Figure \ref{090902BB}. Numerous models have been presented in the literature to explain the spectral features in this burst \citep{2011ApJ...730....1L,2011MNRAS.417.1584B,2012MNRAS.420..468P}.

\noindent{}

\subsubsection{GRB\,090926A}

Detected by GBM \citep{2009GCN..9933....1B} and LAT \citep{2009GCN..9934....1U}, follow-up observations of this GRB were made by Swift-XRT \citep{2009GCN..9936....1V} and Swift-UVOT \citep{2009GCN..9938....1G}. The prompt emission was also observed by Suzaku WAM \citep{2009GCN..9951....1N} and Konus-Wind \citep{2009GCN..9959....1G}. A redshift of z = 2.1062 was obtained by the Very Large Telescope (VLT) X-Shooter spectrograph \citep{2009GCN..9942....1M}. Skynet/PROMPT also observed the optical afterglow \citep{2009GCN..9937....1H}. 

\begin{figure*}
\centering
		\begin{subfigure}[]
                \centering
                \includegraphics[width = 6.1cm, angle=90]{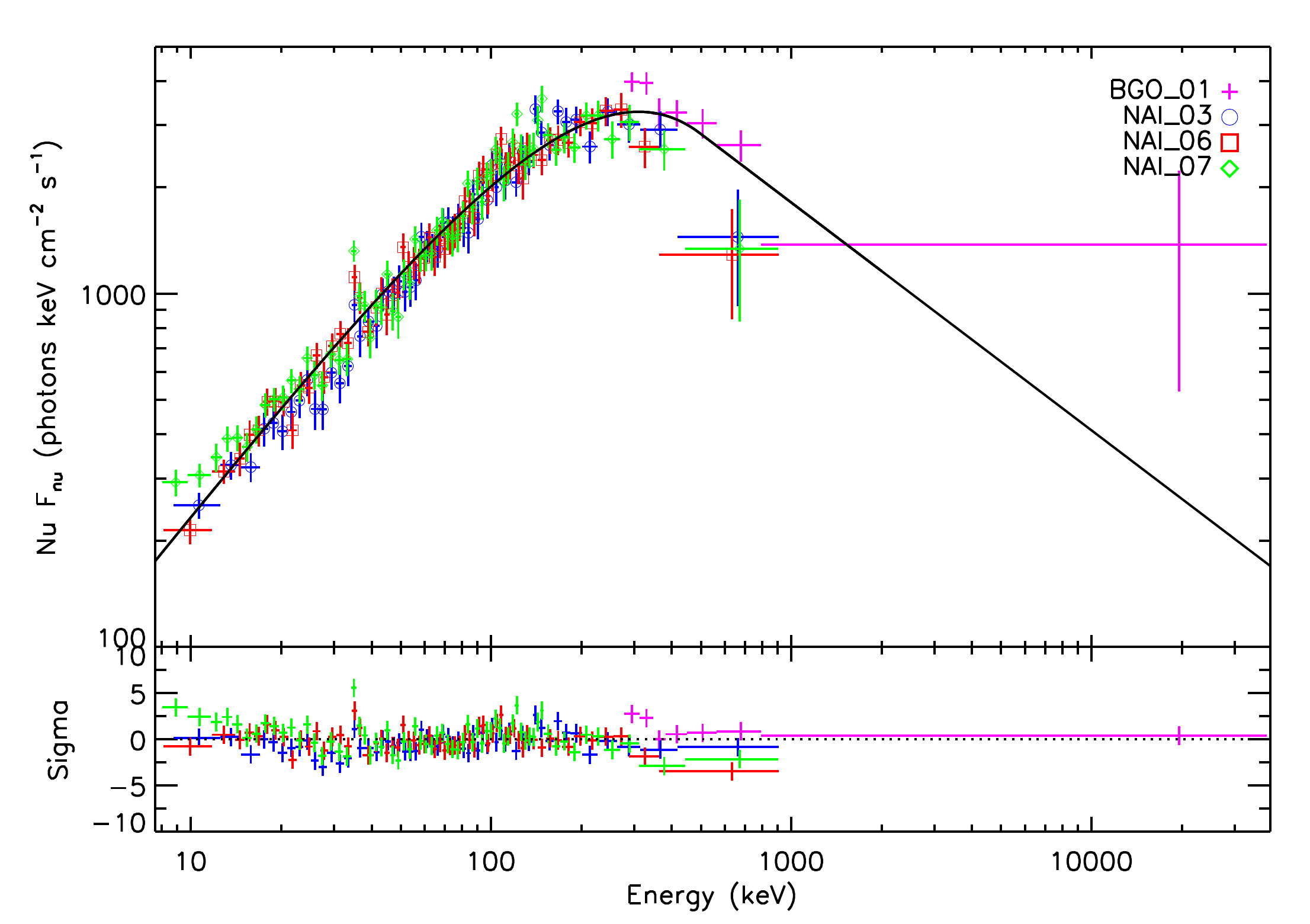}
        \end{subfigure}%
		\begin{subfigure}[]
                \centering
                \includegraphics[width = 6.1cm, angle=90]{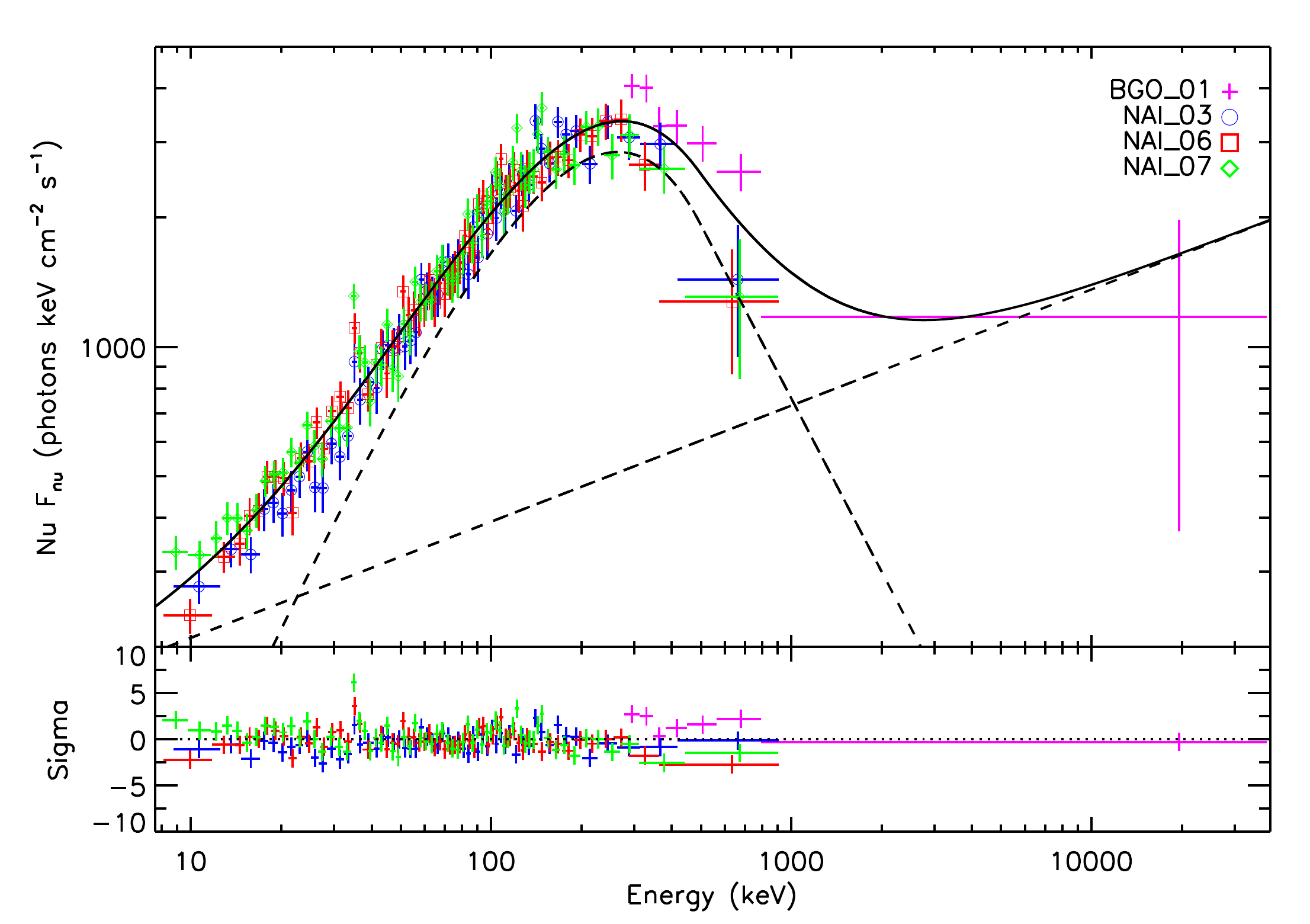}
        \end{subfigure}%
		\begin{subfigure}[]
                \centering
                \includegraphics[width = 6.1cm, angle=90]{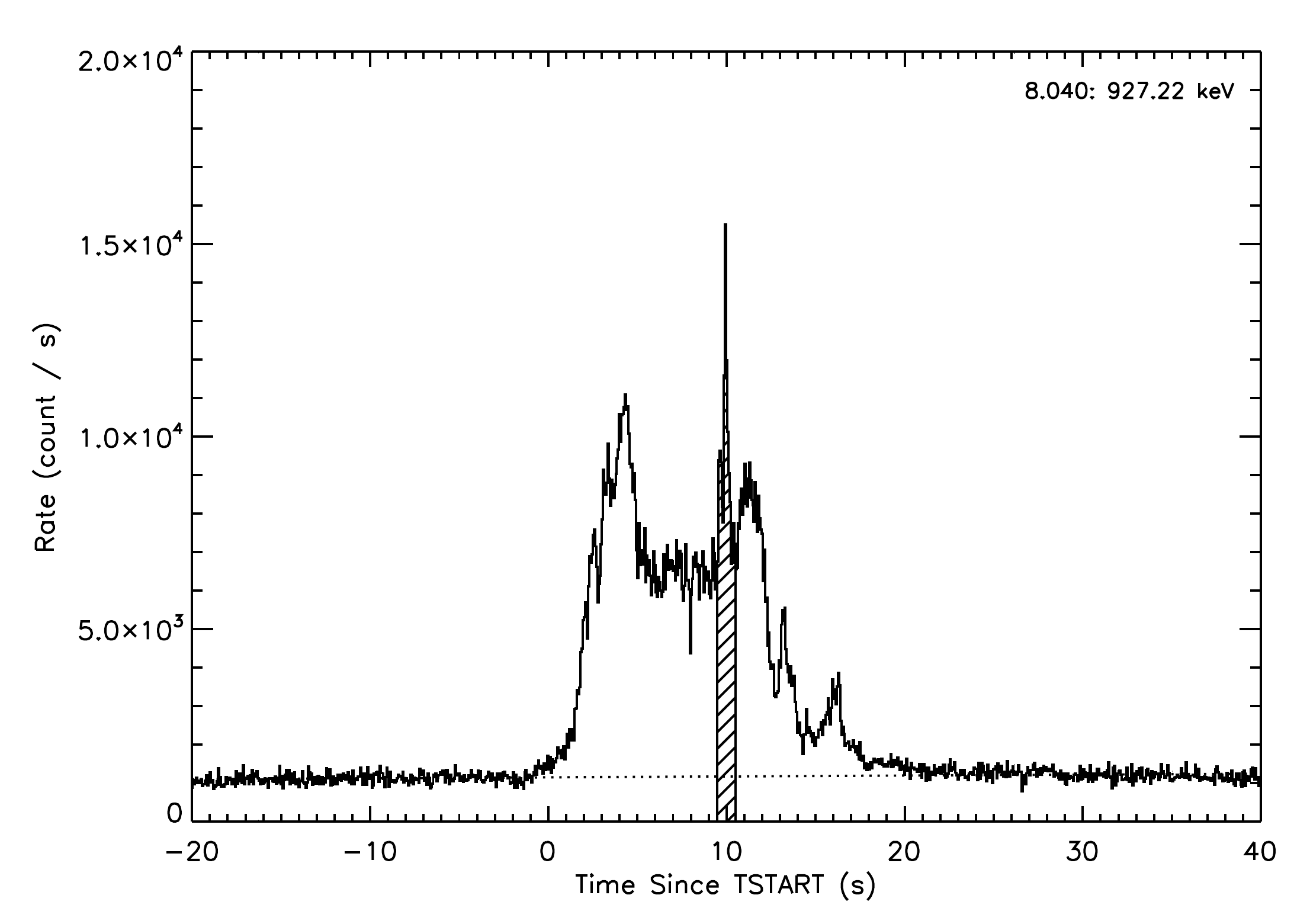}
        \end{subfigure}%
\caption{Spectral fitting of GRB\,090926A. The data were fit with \textbf{a)} a Band function and \textbf{b)} a Band + power-law model. A high-resolution lightcurve is displayed in \textbf{c)} to highlight the short sharp spike in the data.}
\label{090926}
\end{figure*}

GRB\,090926A has also been observed in joint GBM and LAT spectral fitting to necessitate an additional power-law component over a wide spectral range \citep{2011ApJ...729..114A}. Thirteen time bins were analysed between 0 - 17 s with one time bin showing a significant excess. The excess is observed during a short sharp spike in the lightcurve which is shown in Figure \ref{090926}. When an additional power-law is fit with a Band function, the fit statistic shows a large improvement. The additional power-law index  $\alpha$ = -1.73$^{+0.04}_{-0.04}$, obtained by using GBM data only, is again consistent with the power-law observed in the LAT at high energies $\alpha$ = -1.71$^{+0.02}_{-0.05}$ \citep{2011ApJ...729..114A}. 

\noindent{}
 
\section{Discussion and Conclusions}
Although an empirical model, a Band function has been used to model the majority of GRB spectra since it was first proposed in 1993 \citep{1993ApJ...413..281B}. However recent GRB studies with \textit{Fermi} have suggested that there are a number of GRBs whose spectral behaviour differs from a Band function, with a number of different spectral components invoked in order to explain these observations \citep[e.g.][]{2009ApJ...705L.191A, 2011ApJ...741...24B, 2011ApJ...727L..33G, 2012MNRAS.420..468P}. The \textit{Fermi} data cover a wide energy band with high spectral resolution and so the systematic study of low-energy spectral deviations in bright \textit{Fermi} bursts presented here enables the frequency of such events to be investigated. Spectral deviations to a Band function occurring throughout the GBM energy range also affect the spectrum at low energies, and so the current method provides a systematic study over the full GBM energy range, assuming that the spectral data are adequately modelled by a Band function.  

GBM observes 3 out of 45 ($\sim$ 7\%) GRBs with spectral deviations present in the time-integrated data. The probability that our observed rate is consistent with the BATSE observed rate is $\sim$ 7.0\% or within 1.8 $\sigma$. 
An excess is observed in one GRB (GRB\,090902B) and deficits are observed in two (GRB\,090424 and GRB\,081215A) in the time-integrated data, while BATSE observed only excesses at low energies. There are a number of possible causes for the differences between the results obtained from the two instruments. As described in $\S$~\ref{GBMBATSE}, the study of low-energy GRB spectra with BATSE was performed using two data types from two different SDs, with the observations depending on a single DISCSP1 data point at low energies. The structure of these datatypes mean that the low and high-energy bounds of the DISCSP1 point can vary from one GRB to another. GBM has the advantage of a single data type with high spectral resolution over its full energy range allowing deviations to be investigated in fixed energy ranges. Multiple GBM detectors were also used for this analysis, with GRBs having only one `good' detector being analysed for 3 out of 45 GRBs. The effective areas of both instruments are also different. The DRMs of the GBM GRBs are well defined due to their small source location uncertainties, with 80 \% of the GRBs with spectral deviations having \textit{Swift}-XRT locations. It is therefore difficult to make a direct comparison between the GBM and BATSE results. For the reasons outlined above we consider that the GBM result is more robust. 

Since spectral evolution throughout the duration of a GRB significantly affects the spectral shape, a time-resolved analysis was performed for all of the GRBs in the sample. Spectral deviations are found in 11\% of the GRBs in at least one time interval by applying the methods outlined in this paper. Our work shows that the features in the spectra are time-dependent and a time-resolved analysis is necessary to limit the effects of averaging multiple spectra. It is interesting to note that 4 of the 5 GRBs with significant features in the time-resolved analysis are in the top 5 most fluent GRBs in the sample. A bias exists whereby additional models are more likely to be required in high fluence GRBs since these GRBs have enough statistics to confidently define an additional component, if present.

The observed deviations may be due to significant spectral evolution in the burst (or time interval) or additional spectral features/components in the spectrum. In the time-integrated analysis, the deviation found in GRB\,081215A may be due to spectral evolution, while the deviations in GRB\,090424 and GRB\,090902B could be explained by additional spectral features. For those GRBs with deviations found in shorter time intervals (GRB\,090902B and GRB\,090926A), a power-law improves the fit at lower energies which is consistent with the high-energy observations in the LAT data \citep{2009ApJ...706L.138A, 2011ApJ...729..114A}. For GRB\,090424 and GRB\,090820 an additional blackbody  component provides a significantly better fit to the data than an additional power-law, showing that a number of different spectral components may be present in GRBs. GRB\,090323 is adequately fit invoking either of the two additional components. However as there is little evidence for LAT emission \citep{2011AIPC.1358...47P}, it is unlikely to be due to an additional power-law component extending to high energies and a spectral cut-off between the GBM and LAT energy bands is possible. 

The advantage of the method presented in this paper is that it can demonstrate the requirement for an additional spectral component with respect to a Band function without any prior knowledge of the nature of that extra component. Theoretical models of the GRB emission process must be able to explain why the majority of GRBs are adequately modelled by a Band function and also explain the observed deviations. Many physical models have recently been proposed (see $\S$~\ref{intro}), however in many cases the spectral data do not have adequate statistics for one model to be deemed more statistically significant than another. 

This technique demonstrates a systematic method to search GRBs for additional components at low-energies. The method finds several GRBs in which features have previously been noted as well as discovering deviations in GRB\,090424 and GRB\,090323. Deviations are found in a large fraction of high fluence GRBs; this method may be unable to find low-energy deviations in fainter GRBs. This suggests that deviations from a Band function may be common in all GRBs but the statistics do not allow us to test this hypothesis.

\begin{acknowledgements}
      DT acknowledges support from Science Foundation Ireland under grant number 09-RFP-AST-2400. GF acknowledges the support of the Irish Research Council. SF acknowledges the support of the Irish Research Council for Science, Engineering and Technology, cofunded by Marie Curie Actions under FP7. SG was supported by the NASA Postdoctoral Program (NPP) at the NASA/Goddard Space Flight Center, administered by Oak Ridge Associated Universities through a contract with NASA. SG acknowledges financial support through the Cycle-4 NASA Fermi Guest Investigator program. AvK acknowledges support by the German Aerospace Center, through DLR 50 QV 0301.
\end{acknowledgements}

\bibliographystyle{aa}
\bibliography{referencelib_hacked,mnemonic}
%\begin{thebibliography}{}

%\end{thebibliography}

\end{document}